\documentclass[runningheads]{svmult}

\usepackage{makeidx}   
\usepackage{graphicx}  
\usepackage{subeqnar}  
\usepackage{multicol}  
\usepackage{physprbb}  
\makeindex             

\def\ave#1{\langle #1 \rangle}

\def\rund#1{\left( #1 \right)}
\def \la {\mathrel{\vcenter
     {\offinterlineskip \hbox{$<$}\hbox{$\sim$}}}}
\def \ga {\mathrel{\vcenter
     {\offinterlineskip \hbox{$>$}\hbox{$\sim$}}}}
\begin{document}
\title*{Supernova Explosions and Neutron Star Formation}
\toctitle{Supernova Explosions and Neutron Star Formation}
%
%
\titlerunning{Supernova Explosions and Neutron Star Formation}
%
\author{Hans-Thomas Janka, Konstantinos Kifonidis, 
\and Markus Rampp}
\authorrunning{H.-Th. Janka, K. Kifonidis, and M. Rampp}
%
%
\institute{Max-Planck-Institut f\"ur Astrophysik,
           Karl-Schwarzschild-Str.\ 1, D-85741 Garching, Germany}

\maketitle              

\begin{abstract}
The current picture of the collapse and explosion of massive stars
and the formation of neutron stars is reviewed. According to the 
favored scenario, however by no means proven and undisputed, neutrinos
deposit the energy of the explosion in the stellar medium which 
surrounds the nascent
neutron star. Observations, in particular of Supernova~1987A, suggest
that mixing processes play an important role in the expanding star,
and multi-dimensional simulations show that these are linked to 
convective instabilities in the immediate vicinity of the neutron star.
Convectively enhanced energy transport inside the neutron star can
have important consequences for the neutrino emission and thus the
neutrino-heating mechanism. This also holds for a suppression of the neutrino
interactions at nuclear densities. Multi-dimensional hydrodynamics, general
relativity, and a better understanding of the neutrino interactions in 
neutron star matter may be crucial to resolve the problem that state-of-the-art
spherical models do not yield explosions even with a very accurate treatment
of neutrino transport by solving the Boltzmann equation.
\end{abstract}

\section{Introduction}

Baade and Zwicky~\cite{JKRref:baazwi} were the first who speculated about a 
connection between
supernova explosions and the origin of neutron stars. They recognized that stellar
cores must become unstable because neutrons, being produced by captures of 
degenerate electrons on protons, define an energetically advantageous state. The
gravitational binding energy liberated by the collapse of a stellar core
could power a supernova explosion. More than thirty years later, Colgate and
White explored this idea by performing numerical simulations~\cite{JKRref:colwhi}. 
Their models 
showed that the prompt hydrodynamical shock, which forms at the moment of core 
bounce, is not able to reach the outer layers of the star due to severe energy 
losses by photointegration of iron nuclei. Therefore they suggested
that the neutrinos, which are emitted in huge numbers from the hot, collapsed
core and carry away the binding energy of the forming neutron star,
deposit a fraction of their energy in the stellar medium external to
the shock. Later, more accurate simulations with an improved treatment 
of the input physics such as nuclear equation of state and neutrino 
transport~\cite{JKRref:bru85,JKRref:bru87,JKRref:bru89,JKRref:bru93,JKRref:myrblu,JKRref:barcoo,JKRref:swelat}
confirmed the failure of the prompt shock, but could not find the neutrino-driven
explosion imagined by Colgate and White.

The discovery of weak neutral currents, whose existence has been
predicted by the standard model of electro-weak interactions, had a major 
influence also on supernova models. It became clear that the neutrinos 
are strongly coupled to the stellar plasma and can escape only on timescales
of seconds. This implied that the neutrino luminosities did not become large
enough that the rate of energy or momentum transfer by neutrino interactions 
with the nuclei in the stellar
plasma ahead of the supernova shock could become dynamically important.  
In 1982, however, Jim Wilson discovered~\cite{JKRref:wil} that the stalled prompt
shock can be revived by neutrino heating in the dissociated postshock medium
on a longer timescale. While immediately after shock breakout neutrino emission
extracts energy from the shocked gas, energy transfer from neutrinos to the 
postshock medium is favored hundreds of milliseconds
later. If the shock has expanded to a larger radius and the postshock 
temperature has decreased, energetic neutrinos, which stream up from deeper
regions, can deposit a small fraction of their energy in a gain layer behind the
shock~\cite{JKRref:betwil}. Although the efficiency is much lower than imagined 
by Colgate
and White --- typically only a few per cent of the neutrino energy are left in
the stellar medium --- the neutrino heating mechanism can yield the energy of
about $10^{51}\,$erg for the explosion of a massive star like, e.g., 
Supernova~1987A.

Although Wilson's simulations showed the principle viability of this
mechanism and theoretical investigations enlightened the underlying
physics~\cite{JKRref:bet90,JKRref:bet93,JKRref:bet95,JKRref:bet97}, 
later simulations revealed a strong sensitivity 
to the details of the post-bounce evolution of the collapsed stellar 
core~\cite{JKRref:wietal,JKRref:barcoo,JKRref:jamu96,JKRref:mez98b}. 
General relativity, the nuclear equation of state and 
corresponding properties of the nascent neutron star, the treatment of the
neutrino transport and neutrino-matter interactions, and the structure of the 
collapsing star have important influence. In particular, later models
of Mayle and Wilson~\cite{JKRref:wima88,JKRref:wima93} produced sufficiently 
powerful explosions only 
in case of enhanced neutrino emission from the nascent neutron star due to
neutron-finger instabilities. Whether such mixing processes take place, however,
is not clear to date~\cite{JKRref:brmedi}. Alternatively, the hot neutron star might
develop Ledoux convection~\cite{JKRref:epst,JKRref:burlat} 
due to negative gradients of lepton number
and/or entropy, a possibility which is suggested by the structure obtained 
in state-of-the-art spherical models for the neutrino-cooling phase of
proto-neutron stars~\cite{JKRref:pons,JKRref:mirpon}
and by two-dimensional hydrodynamical simulations~\cite{JKRref:kejamu},
but is not generally accepted~\cite{JKRref:mez98a}.

Supernova~1987A has provided us with evidence for large-scale mixing processes,
which involve the layers in the near vicinity of the neutron star where 
radioactive nuclei are produced. Indeed, multi-dimensional simulations have shown the 
existence of hydrodynamical instabilities in the neutrino-heated region already 
during the first second of the 
explosion~\cite{JKRref:heetal,JKRref:milwil,JKRref:jamu95,JKRref:jamu96,JKRref:buetal}.
This convective overturn behind the shock was recognized to allow for an explosion
even when spherical models do not explode. Also, it enhances the explosion 
energy and sets the timescale until the shock gains momentum.
But still it is not clear whether it is sufficient and crucial for the 
neutrino-driven mechanism to be at work, or whether it is just an 
unavoidable side-effect~\cite{JKRref:mez98b}.
The existence of these mixing processes, however, definitely means that
the events leading to an explosion, the energetics of the explosion, the 
nucleosynthesis of radioactive elements during the first second, and the 
composition and distribution of the ejecta can be understood 
quantitatively only by multi-dimensional simulations.

Unfortunately, the 24 neutrinos from Supernova~1987A, which were recorded
in the underground experiments of Kamioka, IMB and Baksan~\cite{JKRref:KIIIMB}, 
did not provide enough statistical information to draw conclusions on the events
which instigated the explosion of the star. Although these neutrinos confirm
the basic picture of stellar core collapse and neutron star formation, they
are not suitable to support the neutrino-driven mechanism. Therefore the 
latter may be considered as the currently most favorable explanation for 
the explosion, but empirical evidence cannot be put forward and numerical
models do not draw a clear and unambiguous picture. 

It cannot be excluded that
the energy for supernova explosions of massive stars is provided by some other
mechanism than neutrino heating, for example by magneto-hydrodynamical processes.
However, we know that neutrinos with the expected characteristics are emitted 
from collapsed stellar cores,
we know that these neutrinos carry away the gravitational binding energy of
the nascent neutron star, we know that neutrino heating {\em must} occur 
behind the stalled shock some time after core bounce, we know that analytic
studies and numerical simulations find explosions for a suitable combination 
of conditions, we know that the energy transfer by neutrinos can be strong 
enough to account for ``normal'' explosions with a canonical energy of 
$\sim 10^{51}\,$erg, and we know that the timescale of shock rejuvenation by
neutrino heating determines the mass cut such that accretion of the collapsed
stellar core and later fallback lead to neutron star masses and supernova
nucleosynthesis in rough agreement with 
observations~\cite{JKRref:fryer,JKRref:frykal,JKRref:woowea}.
Of course, the discrepant results of numerical simulations are unsatisfactory, 
and major problems are nagging: Why do the supposedly best and most 
advanced spherical models not produce explosions? Can one trust current
multi-dimensional simulations with their greatly simplified and approximate 
treatment of neutrino transport? Is rotation in neutrino-driven explosions
sufficient to explain the observed
large asphericities and anisotropies in many supernovae~\cite{JKRref:fryheg}?
What is the reason for the kicks by which pulsars are accelerated to average
velocities of several hundred km/s presumably during the supernova explosion?
What powers hyperenergetic supernovae, which seem to release up to 50 times
more kinetic energy than ordinary explosions of massive stars~\cite{JKRref:hyper}?

This paper gives an overview over the major lines of research on the standard 
explosion scenario of massive stars, and on neutron star formation, where 
progress has been achieved during the past few years or will be coming up in the
near future. In Section~2 the physics of neutrino-driven explosions will
be discussed in some detail, and an analytic toy model will be described 
which allows one a deeper understanding of the requirements of neutrino-driven
explosions. In Section~3 the current status of supernova modeling in spherical 
symmetry will be outlined, with particular focus on results with neutrino
transport by solving the Boltzmann equation. In Section~4 the relevance
of convective overturn in the neutrino-heated layer will be discussed. New
results of two-dimensional simulations will be presented, which take into account
the nucleosynthesis during the explosion and follow the shock from the moment of
its formation until it breaks out of the surface of the star. In Section~5, 
two-dimensional hydrodynamical models of the neutrino-cooling phase of newly
formed neutron stars will be described. The effects of rotation and of a 
suppression of neutrino-nucleon interactions by nucleon correlations in the 
nuclear medium will be addressed. Section~6 will contain a summary and conclusions.

\begin{figure}[t]
\begin{center}
\vspace{3truecm}
\end{center}
\caption[]{
Sketch of the post-collapse stellar core during the neutrino heating
and shock revival phase. $R_{\nu}$ is the neutrinosphere radius,
$R_{\rm ns}$ the protoneutron star radius, $R_{\rm g}$ the gain
radius outside of which net neutrino heating exceeds neutrino
cooling, and $R_{\rm s}$ is the shock radius.
The shock expansion is impeded by mass infall
at a rate $\dot M$, but supported by convective energy transport
from the region of strongest neutrino heating into the post-shock
layer. Convection inside the nascent neutron star raises the neutrino
luminosities.
}
\label{JKR:fig-1}
\end{figure}

\section{The Explosion Mechanism}

The physics of neutrino-driven explosions is discussed, first on
the level of basic considerations, then with the help of an
analytic toy model, which allows one to study the competing 
effects that determine the destiny of the stalled supernova shock.

\subsection{Neutrino Heating}

Figure~\ref{JKR:fig-1} displays a sketch of the neutrino cooling and heating
regions outside the proto-neutron star at the center. The main processes
of neutrino energy deposition are the charged-current reactions
$\nu_e+n \rightarrow p + e^-$ and $\bar\nu_e+p \rightarrow 
n + e^+$~\cite{JKRref:betwil}.
With the neutrino luminosity, $L$, the average squared neutrino energy, 
$\langle\epsilon^2\rangle$, and the mean value of the cosine of the angle
$\theta_{\nu}$ of the
direction of neutrino propagation relative to the radial direction,
$\ave{\mu} = \ave{\cos\theta_{\nu}}$, being defined as
moments of the neutrino phase space distribution function
$f(r,t,\mu,\epsilon)$ by integration over energies $\epsilon$ and angles
$\mu$ according to
\begin{eqnarray}
L&=& 4\pi r^2\,{2\pi c\over (hc)^3}\,\int_{-1}^{+1}{\rm d}\mu
\int_0^\infty{\rm d}\epsilon\,\epsilon^3\mu f \ ,\\
\ave{\epsilon^2}&=&\int_{-1}^{+1}{\rm d}\mu\int_0^\infty{\rm d}\epsilon\,
\epsilon^5 f\cdot \left\{\int_{-1}^{+1}{\rm d}\mu
\int_0^\infty{\rm d}\epsilon\,\epsilon^3 f \right\}^{-1} \ ,\\
\ave{\mu}&=&\int_{-1}^{+1}{\rm d}\mu\int_0^\infty{\rm d}\epsilon\,
\epsilon^3\mu f \cdot \left\{ \int_{-1}^{+1}{\rm d}\mu
\int_0^\infty{\rm d}\epsilon\,\epsilon^3 f \right\}^{-1} \ ,
\end{eqnarray}
the heating rate per nucleon ($N$) is approximately given by
\begin{eqnarray}
Q_{\nu}^+ &\approx & 110\cdot \rund{
{L_{\nu_e,52}\langle\epsilon_{\nu_e,15}^2\rangle\over r_7^2\,\,\ave{\mu}_{\nu_e}}
\,Y_n\,+\,
{L_{\bar\nu_e,52}\langle\epsilon_{\bar\nu_e,15}^2\rangle\over r_7^2\,\,
\ave{\mu}_{\bar\nu_e}}\,Y_p } 
\quad
\left\lbrack {{\rm MeV}\over {\rm s}\cdot N}\right\rbrack  \nonumber \\
          &\approx & 55\cdot {L_{\nu,52}\langle\epsilon_{\nu,15}^2\rangle
\over r_7^2 \,\, f}
\quad \left\lbrack {{\rm MeV}\over {\rm s}\cdot N}\right\rbrack \; ,
\label{JKR:eq-1}
\end{eqnarray}
where $Y_n$ and $Y_p$ are the number fractions of free neutrons and protons
(number densities divided by baryon number density), respectively. In the
second equation $Y_n + Y_p \approx 1$, and equal luminosities and
spectra for $\nu_e$ and $\bar\nu_e$ were assumed.
$L_{\nu,52}$ denotes the total luminosity of $\nu_e$ plus $\bar\nu_e$ in
$10^{52}\,{\rm erg/s}$, $r_7$ the radial position in $10^7\,{\rm cm}$,
$\langle\epsilon_{\nu,15}^2\rangle$ is measured in units of $15\,{\rm MeV}$,
and $f = \ave{\mu}_{\nu}$ is very small in the opaque regime where
the neutrinos are isotropic, adopts a value of about 0.25 around the
neutrinosphere, and approaches unity for radially streaming neutrinos
very far out. Note that the ``flux factors'' $\ave{\mu}_{\nu}$ determines
the neutrino energy density at a radius $r$ according to
$\varepsilon_{\nu}(r) = L_{\nu}/(4\pi r^2c \ave{\mu}_{\nu})$.

Using this energy deposition rate, neglecting energy losses due to the
re-emission of neutrinos, and assuming that the gravitational binding energy
of a nucleon in the neutron star potential is (roughly) balanced by the sum
of internal and nuclear recombination energies after accretion of the infalling
matter through the shock, one can estimate (very approximately) the explosion
energy to be of the order
\begin{equation}
E_{\rm exp} \,\sim\, 10^{51}\cdot
{L_{\nu,52}\langle\epsilon_{\nu,15}^2\rangle\over R_{\mathrm{g},7}^2\,\, f}\,
\left({\Delta M \over 0.1\,M_\odot}\right)
\left({\Delta t\over 0.1\,{\rm s}}\right)
-\,E_{\rm gb} +\,E_{\rm nuc} \quad \left\lbrack \rm erg\right\rbrack \, .
\label{JKR:eq-2}
\end{equation}
Here $\Delta M$ is the heated mass, $\Delta t$ the typical heating timescale,
$E_{\rm gb}$ the (net) total gravitational binding energy of the overlying,
outward accelerated stellar layers, and $E_{\rm nuc}$ the additional energy
from explosive nucleosynthesis, which is a significant contribution of
a few $10^{50}\,{\rm erg}$ only for progenitors with main sequence masses
above $20\,M_{\odot}$, and which roughly compensates
$E_{\rm gb}$\footnote{The latter statement is supported by the following
argument (S.~Woosley, personal communication): Material with a
specific gravitational binding energy $\Phi_{\rm grav}$ which is equal to
or larger than the nuclear energy release per gram in Si burning,
$e_{\rm nuc}\sim 10^{18}\,$erg/g, is located
interior to the radius where the temperatures can become high enough
($T\ga 5\times 10^9\,$K) for explosive nucleosynthesis of $^{56}$Ni.
>From $\Phi_{\rm grav} = GM/r \ga 10^{18}\,$erg/g one estimates a radius
of $r\la 2\times 10^8\,$cm, and from ${4\over 3}\pi r^3aT^4\sim 10^{51}\,$erg
with $T\ga 5\times 10^9\,$K one finds $r \la 4\times 10^8\,$cm.
This means that the energy release from explosive nucleosynthesis is easily
able to account for the gravitational binding energy of the burning material.}.

It is not easy to infer from Eq.~(\ref{JKR:eq-2}) the dependence of the 
explosion energy on the neutrino luminosity. On the one hand, 
\begin{equation}
E_{\mathrm{exp}}\,\propto\, Q_{\nu}^+V\Delta t\,\propto\,
{L_{\nu}\ave{\epsilon_{\nu}^2}\over R_{\mathrm{g}}^2}\,\Delta M\Delta t
\,\propto\,L_{\nu}\Delta\tau\,\Delta t \, ,
\label{JKR:eq-2a}
\end{equation}
where $V$ is the heated volume between gain radius and shock, and $\Delta\tau$ 
the optical depth of the heating layer. On the other hand,
\begin{equation}
\Delta\tau\,\propto\,\ave{\epsilon_{\nu}^2}R_{\mathrm{g}}\,\rho_{\mathrm{g}}
\,\propto\,\ave{\epsilon_{\nu}^2}R_{\mathrm{g}}T_{\mathrm{g}}^3\,
\propto\,L_{\nu}^{1/2}\ave{\epsilon_{\nu}^2}^{3/2}\, .
\label{JKR:eq-2b}
\end{equation}
Here $\rho\propto T^3$ was assumed for the relation between density and 
temperature in the heating layer~\cite{JKRref:bet97},
and $Q_{\nu}^+(R_{\mathrm{g}}) = Q_{\nu}^-(R_{\mathrm{g}})$ was used at the
gain radius, where neutrino heating is balanced by neutrino cooling. The 
energy loss rate $Q_{\nu}^-$ by neutrinos produced in capture reactions of 
nondegenerate electrons and positrons on nucleons, 
scales with $T^6$. Combining Eqs.~(\ref{JKR:eq-2a}) 
and (\ref{JKR:eq-2b}) yields
\begin{equation}
E_{\mathrm{exp}}\,\propto\,L_{\nu}^{3/2}\ave{\epsilon_{\nu}^2}^{3/2}\Delta t\,
\propto\, L_{\nu}^{9/4}\Delta t\, ,
\label{JKR:eq-2c}
\end{equation}
when $\ave{\epsilon_{\nu}^2}\propto T_{\nu}^2\propto L_{\nu}^{1/2}$ is used
for black-body like emission. 

If the expansion velocity were simply proportional to $E_{\mathrm{exp}}^{1/2}$,
in which case the time $\Delta t$ for the shock to reach a given radius would be
$\Delta t \propto E_{\mathrm{exp}}^{-1/2}$, then 
$E_{\mathrm{exp}}\propto L_{\nu}^{3/2}$~\cite{JKRref:bet97}.
However, when shock expansion sets in, most of the energy is internal energy,
but not kinetic energy, making the relation $\Delta t \propto E_{\mathrm{exp}}^{-1/2}$
very questionable. The actual variation of $\Delta t$ with the inverse of the 
neutrino luminosity can be steeper.

\subsection{Requirements for Neutrino-Driven Explosions}

In order to get explosions by the delayed neutrino-heating mechanism, certain
conditions need to be fulfilled. Expansion of the postshock region requires
sufficiently large pressure gradients near the radius $R_{\rm cut}$ of the
developing mass cut. If one neglects self-gravity of the gas in this region and
assumes the density profile to be a power law, $\rho(r) \propto r^{-n}$
(which is well justified according to numerical simulations
which yield a power law index of $n \approx 3$~\cite{JKRref:bet93}),
one gets $P(r)\propto r^{-n-1}$ for the pressure in an atmosphere
near hydrostatic equilibrium. Outward acceleration
is therefore maintained as long as the following condition for the
``critical'' internal energy density $\varepsilon$ holds:
\begin{equation}
\left. {\varepsilon_{\rm c}\over GM\rho/r}\right|_{R_{\rm cut}} \,>\,
{1\over (n+1)(\gamma-1)}\,\cong\,{3\over 4} \; ,
\label{JKR:eq-3}
\end{equation}
where use was made of the relation $P = (\gamma-1)\varepsilon$.
The numerical value was obtained for $\gamma = 4/3$ and $n = 3$. This
condition can be converted into a criterion for the entropy per baryon,
$s$. Using the thermodynamical relation for the entropy density
normalized to the baryon density $n_b$,
$s = (\varepsilon + P)/(n_b T) - \sum_i \eta_i Y_i$
where $\eta_i$ ($i = n,\,p,\,e^-,\,e^+$) are the particle
chemical potentials divided by the temperature, and assuming
completely disintegrated nuclei behind the shock so that the
number fractions of free protons and neutrons are $Y_p = Y_e$ and
$Y_n = 1 - Y_e$, respectively, one gets
%
\begin{equation}
s_{\rm c}(R_{\rm cut}) \,\ga\,
\left. 14\,\,{M_1\over r_7 \, T}\,\right|_{R_{\rm cut}}\,-\,
\left. \ln\left(1.27\cdot 10^{-3}\,\,{\rho_9\, Y_n\over T^{3/2}}\right)
\right|_{R_{\rm cut}}\quad \left\lbrack k_{\rm B}/N\right\rbrack
\;.
\label{JKR:eq-4}
\end{equation}
In this approximate expression a term with a factor $Y_e$
was dropped (its absolute value being usually less than 0.5 in
the considered region), nucleons were assumed to obey Boltzmann
statistics, and $T$ is
measured in MeV, $M_1$ in units of $M_\odot$, $\rho_9$ in
$10^9\,{\rm g/cm}^3$, and $r_7$ in $10^7\,{\rm cm}$. Inserting
typical numbers ($T\approx 1.5\,{\rm MeV}$, $Y_n \approx 0.3$,
$R_{\rm cut}\approx 1.5\cdot 10^7\,{\rm cm}$), one finds
$s > 15\,k_{\rm B}/N$, which gives
an estimate of the entropy in the heating region when expansion is
going to take place.

Since the entropy and energy density in the postshock later are 
raised by neutrino energy deposition, the conditions of Eqs.~(\ref{JKR:eq-3})
and (\ref{JKR:eq-4}) imply requirements on the neutrino emission of the
proto-neutron star. These can be derived by the following considerations.
A stalled shock is converted into a moving
one only, when the neutrino heating is strong enough to increase the pressure
behind the shock by a sufficient amount. From the Rankine-Hugoniot relations
at the shock, a criterion can be deduced for the heating rate
per unit mass, $q_{\nu}$, behind the shock, which leads to a positive postshock
velocity ($u_1 > 0$)~\cite{JKRref:bru93}:
\begin{equation}
q_{\nu}\,>\,{2\beta - 1\over \beta^3(\beta-1)(\gamma-1)}\,
{|u_0|^3 \over \eta R_{\rm s}} \, .
\label{JKR:eq-5}
\end{equation}
Here $\beta$ is the ratio of postshock to preshock density,
$\beta = \rho_1/\rho_0$, $\gamma$ the adiabatic index of the gas
(assumed to be the same in front and behind the shock), and $\eta$
defines the fraction of the shock radius $R_{\rm s}$ where net heating
by neutrino processes occurs: $\eta = (R_{\rm s}-R_{\rm g})/R_{\rm s}$.
$u_0$ is the preshock velocity, which is a fraction $\alpha$
(analytical and numerical calculations show that typically
$\alpha \approx 1/\sqrt{2}$) of the free fall velocity:
$u_0 = \alpha\sqrt{2GM/r}$. Assuming a strong shock, one has
$\beta = (\gamma+1)/(\gamma-1)$, which becomes $\beta = 7$ for
$\gamma = 4/3$. With typical numbers,
$R_{\rm s} = 100\,{\rm km}$, $M = M_1 = 1\,M_{\odot}$, and 
$\eta\approx 0.4$, one derives for the critical luminosity of
$\nu_e$ plus $\bar\nu_e$:
\begin{equation}
L_{\nu,52}\langle\epsilon_{\nu,15}^2\rangle\ >\ 4.4\,\,
{M_1^{3/2}\over R_{{\rm s},7}^{1/2}} \, .
\label{JKR:eq-6}
\end{equation}
Since this discussion was very approximate, e.g., the reemission of neutrinos 
was ignored and properties depending on the structure of the collapsed stellar
core were absorbed into free parameters, the analysis
cannot yield a quantitatively meaningful
value for the threshold luminosity. However, the existence of a lower bound on
the neutrino luminosity as found in numerical 
simulations~\cite{JKRref:jamu95,JKRref:jamu96},
is confirmed. Above this threshold value, neutrino heating of the gas
behind the shock is strong enough to drive an expansion of the gain layer.

%
%
\begin{figure}[t]
\begin{center}
\includegraphics[width=.475\textwidth]{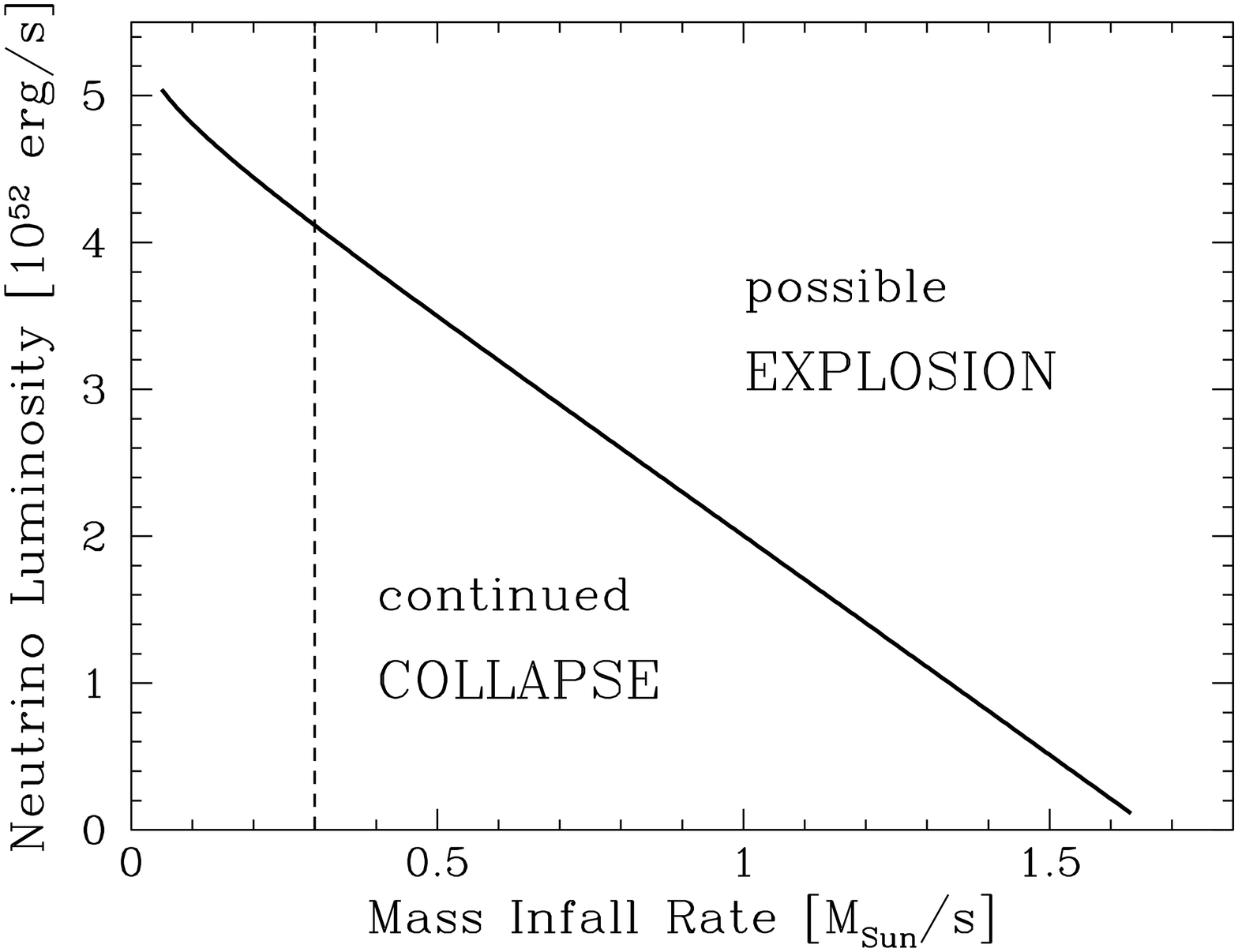}
\hspace{5pt}
\includegraphics[width=.475\textwidth]{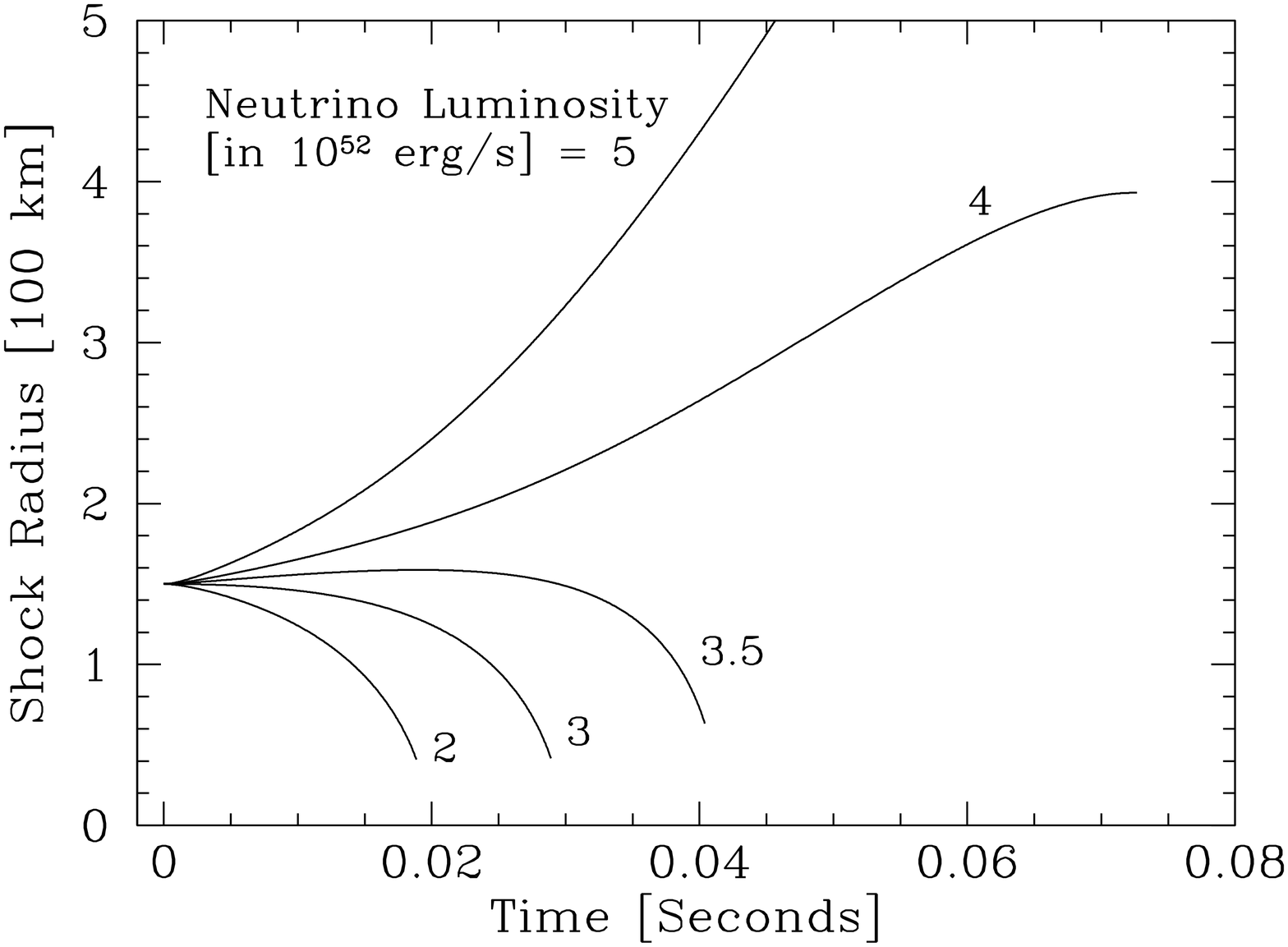}
\end{center}
\caption[]{
{\it Left:}
Phase diagram for successful explosion or continued stellar
collapse in dependence upon the rate at which gas falls into the shock and
upon the $\nu_e$ plus $\bar\nu_e$ luminosity of the nascent neutron star. 
The mass of the neutron star was assumed to be 1.25$\,M_{\odot}$.
{\it Right:} 
Shock positions as functions of time for different values
of the $\nu_e$ plus $\bar\nu_e$ luminosity of the neutron star (according
to the labels) and fixed value of the rate at which gas falls into the shock
(marked by the dashed line in the left plot). For neutrino luminosities
above the critical value (where the dashed line crosses the solid line
in the left plot) explosions can occur.
}
\label{JKR:fig-2}
\end{figure}


\subsection{Analytic Toy Model}

The discussion in the previous section was overly simplified.
The behavior of the stagnant shock does not only depend on the neutrino
heating in the gain region, but is also influenced by the energy loss and
the settling of the cooling layer, and by the mass infall from the 
collapsing star star ahead of the shock. Only hydrodynamical simulations
can determine the shock evolution in response to these different, partly
competing effects. Analytic discussions, however, can help one understanding
the significance of different effects and thus can supplement more detailed,
but less transparent supercomputer calculations.

The conditions in the region of neutrino heating and the details of
the heating process were analysed by Bethe and Wilson~\cite{JKRref:betwil} 
and Bethe~\cite{JKRref:bet90,JKRref:bet93,JKRref:bet95,JKRref:bet97}. 
Burrows and Goshy~\cite{JKRref:burgos} considered the post-bounce accretion phase
of the supernova shock as a quasi steady-state situation, and thus replaced the
time-dependent partial differential equations of hydrodynamics by a set of
ordinary differential equations to determine the radial position of the standing 
shock by an eigenvalue analysis. While this approach captures an interesting 
aspect of the problem, it has serious weaknesses.
The accretion flow between the shock and the neutron star does not need to
be stationary, but neutrino-heated matter may stay in the gain region, or gas
will pile up on the forming neutron star when neutrinos are unable to remove 
energy quickly enough for the gas to settle. In particular, when the shock 
accelerates outward, the transition from accretion to outflow cannot be 
described as a stationary situation.

Since the gas falling into the stalled shock is strongly decelerated and the 
postshock velocities are smaller than the local sound speed, the structure 
of the collapsed stellar core is rather simple and can be well described by 
hydrostatic equilibrium. This allows for an approximate treatment, which
is complementary to the approach taken by Burrows and Goshy~\cite{JKRref:burgos}: 
Integrating the 
stellar structure over radius leads to conservation laws for the total mass
and energy in the gain layer. The time-dependent radius and velocity of the 
shock can then
be obtained as solutions of an {\em initial value problem}, which reflects 
the fact that the destiny of the shock depends on the initial conditions and is
controlled by the cumulative effects of neutrino energy deposition and mass
accumulation in the gain layer~\cite{JKRref:janka}. 

Such an analysis also demonstrates the existence of a threshold value for the neutrino 
luminosity from the neutron star, which is needed to drive shock expansion.
This threshold luminosity depends on the rate, $\dot M$, of mass infall to the 
shock, on the neutron star mass and radius, and to some degree also on the 
shock stagnation radius. Taking into account only the main dependence on $\dot M$,
it can roughly be written as~\cite{JKRref:janka}
\begin{equation}
L_{\nu,{\mathrm{crit}}}(\dot M)\,\approx\,L_0 - L_1\rund{{\dot M\over M_{\odot}/
{\mathrm{s}}}} \, ,
\label{JKR:eq-6a}
\end{equation}
with $L_0 \approx 5\times 10^{52}$~erg$\,$s$^{-1}$ and 
$L_1 \approx 3\times 10^{52}$~erg$\,$s$^{-1}$ for the conditions of
Fig.~\ref{JKR:fig-2}. 

The neutrino heating in the gain layer is not the only important factor
that determines the shock propagation. Energy loss by neutrino emission
in the cooling layer has a considerable influence, because it regulates the
settling of the matter that is accreted by the nascent neutron star, and 
therefore the advection of gas through the heating layer. If cooling is
inefficient, gas piles up on the neutron star and pushes the shock farther 
out. If cooling is very efficient, the gas contracts quickly and more gas is 
dragged downward through the gain radius, extracting mass and energy from
the gain layer and thus weakening the support for the shock. This also means
that the infall velocity behind the shock increases and the timescale for the
gas to stay in the gain layer is reduced. Therefore the efficiency of neutrino
energy deposition drops. Such an effect is harmful for shock expansion. It
can be diminished by higher $\nu_e$ and $\bar\nu_e$ luminosities from the
neutrinosphere, which lead to an enhancement of neutrino absorption relative
to neutrino emission. On the other hand, muon and tau neutrino and antineutrino
production in the accretion layer of the neutron star has a desastrous consequence
for the shock, because it is a sink of energy that leaves the star without
any significant positive effect above the neutrinosphere, where only $\nu_e$ 
and $\bar\nu_e$ can be absorbed by free nucleons.

These different, competing processes combined explain the slope of
the critical line in the left plot of Fig.~\ref{JKR:fig-2} and in 
Eq.~(\ref{JKR:eq-6a}). Shock expansion
and acceleration are easier for high mass infall rates, $\dot M$, into the shock
and for high $\nu_e$ and $\bar\nu_e$ luminosities from the nascent neutron star. 
These luminosities need to be larger when $\dot M$ is small. It must be pointed
out here, however, that this dependence is a consequence of the fact that the 
temperature in the cooling layer is considered as a parameter of the discussion.
It is assumed to be equal to the neutrinospheric temperature and thus
to be mainly determined by the interaction with the neutrino flux from the
core of the neutron star, but not by the mass infall and the dynamics in the 
accretion layer.

Emission of muon and taun neutrinos and antineutrinos from the cooling region
is not included in the results displayed in Fig.~\ref{JKR:fig-2}. It would
move the critical line to appreciably higher values of the $\nu_e$ plus
$\bar\nu_e$ luminosity of the collapsed core, which is given along the
ordinate. 

Neutrino heating is stronger close to the gain
radius than right behind the shock. Using an
isentropic profile in the gain layer, the evaluation, however,
implies very efficient energy transport, e.g., by convective motions
in the gain layer. This enhances the postshock
pressure and reduces the loss of energy from the gain layer, which is associated
with the inward advection of neutrino-heated gas. 

Solutions of the toy model for varied parameters show that the energy in the 
gain layer and therefore the explosion energy of the supernova is limited to
some $10^{51}\,$erg. The reason for this is the following. Neutrino energy 
deposition proceeds by $\nu_e$ and $\bar\nu_e$ absorption on nucleons. The
heated gas expands away from the region of strongest heating as soon as the
nucleons have absorbed an energy roughly of the order of the gravitational
potential energy, with only a small time lag because of the inertia of the 
shock, which is confined by the ram pressure of the collapsing stellar material. 
This does not allow the net energy of the heated gas to become very large.
Typically it is of the order of $\sim 5\,$MeV per nucleon. With a mass in the gain
layer of several 0.01$\,M_{\odot}$ up to $\sim 0.1\,M_{\odot}$, the total 
energy does therefore not exceed a few $10^{51}\,$erg.

%
%
\begin{figure}[t]
\begin{center}
\includegraphics[width=.75\textwidth]{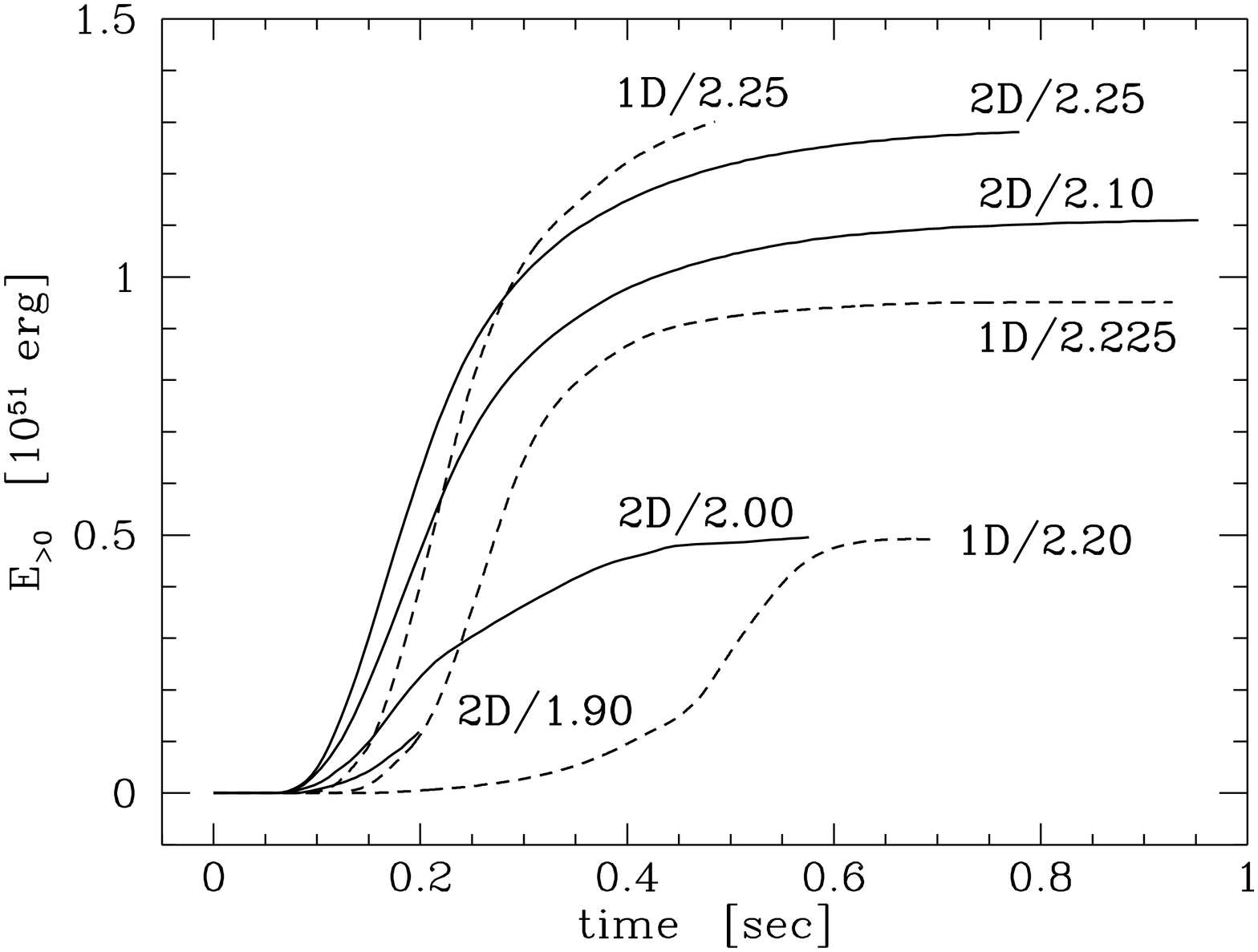}
\end{center}
\caption[]{
Explosion energies $E_{>0}(t)$ for spherically symmetric models
(``1D'', dashed lines) and two-dimensional (``2D'') supernova 
models (solid lines)~\cite{JKRref:jamu95,JKRref:jamu96}. The latter
take into account convective overturn between the supernova shock
and the neutrino-heating region.
The curves display the evolution as a function of time after shock
formation for different values of the $\nu_e$ luminosity
(labeled in units of $10^{52}\,$erg/s). The latter was used as a free parameter
at the surface of the nascent
neutron star and was roughly equal to the $\bar\nu_e$ luminosity.
Below the smallest given luminosities, the
considered $15\,M_{\odot}$ star does not explode in 1D and acquires too low
an expansion energy in 2D to unbind the whole stellar mantle and envelope.
}
\label{JKR:fig-3}
\end{figure}
%
%

Figure~\ref{JKR:fig-3} clearly shows this saturation of the explosion
energy, which occurs when the gain layer and the shock are expanding. In this
sense, neutrino-driven explosions are ``self-regulated'': Further energy
deposition is quenched as the baryons move out of the region of high
neutrino fluxes.

The heating rate increases with the neutrino luminosity and the deposited
energy is higher for a larger mass in the gain layer (Eq.~\ref{JKR:eq-2}).
However, the expansion timescale during which the gas is exposed to high
neutrino fluxes, drops when the heating is stronger. Therefore the
explosion energy is extremely sensitive to the neutrino luminosity only
around the threshold value for getting an explosion. 

Neutrino-driven explosions are likely to be ``delayed'' 
(up to a few 100~ms after core bounce) rather than ``late''
(after a few seconds). The density between the gain radius
and the shock decreases with time because the proto-neutron star
contracts and the gas infall to the shock drops rapidly as time goes on.
Therefore the mass $\Delta M$ in the heated region shrinks
and the shock must recede to a small radius around or even below 
100~km, not favoring a later explosion.

Fulfilling the ``explosion criterion'' of Fig.~\ref{JKR:fig-2} once is
no guarantee for a successful explosion. The push from the heating region
has to be maintained until the material that carries the bulk of the 
energy is moving ballistically, and a fair fraction of this energy has
been converted from internal to kinetic energy. Otherwise, if energy
losses by neutrino emission or $P{\mathrm{d}}V$ on the contracting 
proto-neutron star start to dominate the energy input by neutrino absorption,
the pressure-supported expansion can break down again and re-collapse
can occur. However, once shock expansion sets in, the conditions for 
further neutrino heating improve rapidly, and the optical depth of the
growing gain layer to neutrinos increases. Provided the neutrino 
luminosity does not drop, an explosion becomes unavoidable. This 
requirement favors a high core neutrino luminosity over accretion 
luminosity to power neutrino-driven explosions.

%
%
\begin{figure}[t]
\begin{center}
\vspace{5truecm}
\end{center}
\caption[]{
Trajectories of mass shells in the core of a collapsing 15$\,M_{\odot}$ star
from a Newtonian simulation with Boltzmann neutrino 
transport~\cite{JKRref:rampp,JKRref:ramjan}.
The shells are equidistantly spaced in steps of $0.02\,M_{\odot}$. The 
boundaries of the iron core, silicon shell and neon-magnesium shell are
indicated by bold lines. The fat, solid curve rising up at 0.21 seconds after 
the start of the simulation marks the position of the supernova shock, the dashed
line denotes the gain radius. In this spherically symmetric simulation, 
muon and tau neutrinos
and antineutrinos were neglected, which favors an explosion. Nevertheless,
the shock recedes after having expanded to more than 300$\,$km. 
}
\label{JKR:fig-3a}
\end{figure}
%
%

\section{Status of Spherical Simulations}

Recently, a major shortcoming of previous supernova models has been removed,
at least in spherical models. Instead of treating multi-frequency neutrino 
transport by a flux-limited diffusion 
approximation~\cite{JKRref:wima88,JKRref:wima93,JKRref:myrblu,JKRref:bru85},
the Boltzmann equation can now be solved in connection with hydrodynamical
simulations, either by direct discretisation~\cite{JKRref:mez01}
or by a variable Eddington factor technique~\cite{JKRref:ramjan},
even in the general relativistic case~\cite{JKRref:lie01}. 
For the first time, the numerical deficiencies of the models are therefore 
smaller than the uncertainties of the input physics.

The more accurate treatment of the transport, in particular in the 
semi-transparent neutrino-decoupling region around and outside of the
neutrinosphere up to the shock, favors higher energy transfer to the 
stellar gas in the cooling layer and in the gain 
layer~\cite{JKRref:meetal,JKRref:yaetal}.
Nevertheless, the results of these simulations are disappointing.
In spherical symmetry, the models do not explode, neither in the 
Newtonian (Fig.~\ref{JKR:fig-3a}), nor in the general relativistic case.

These current models, however, neglect convective effects inside the nascent
neutron star as well as in the neutrino-heated region. Since convection has
been recognized to be very important, such simulations do not treat the 
full supernova problem and do not really allow for conclusions about the
viability of the neutrino-driven mechanism. 
Multi-dimensional simulations with Boltzmann neutrino transport are called for.

%
%
\begin{figure}[t]
\begin{center}
\vspace{5truecm}
\end{center}
\caption[]{
Inhomogeneous distribution of neutrino-heated, hot gas which rises
in mushroom-like bubbles, and cooler gas that is accreted through
the supernova shock (bumpy discontinuity at about 3000$\,$km) and
falls in long, narrows streams towards the newly formed neutron 
star at the center. The figure shows a snapshot of the entropy at
300$\,$ms after core bounce and shock formation for a two-dimensional
simulation of a 15$\,M_{\odot}$ star. 
The star was exploded by the neutrino-heating
mechanism by chosing a suitable value 
of the neutrino luminosity from the nascent neutron star, which was 
replaced by an inner boundary condition~\cite{JKRref:kifon,JKRref:kietal}.
The white line encompasses the region where radioactive nickel has 
been formed by nuclear burning in the shock-heated Si layer.
}
\label{JKR:fig-4}
\end{figure}
%
%

\section{Hydrodynamic Instabilities during the Explosion}

During the explosion of a supernova, hydrodynamic instabilities and convective
processes can occur on different scales in space and time. Convective motions
inside the nascent neutron star can speed up the energy transport 
and raise the neutrino luminosities during a period of seconds (Section~5).
In the neutrino-heated region, convective overturn during the first second of
the explosion carries hot matter towards
the shock front and brings cool gas into the region of strongest neutrino heating
near the gain radius. This has important influence on the start of the explosion
and the nucleosynthesis of radioactive elements. When the shock propagates 
through the mantle and envelope of the disrupted star, Rayleigh-Taylor
instabilities destroy the onion-shell structure of the progenitor and mix 
radioactive material with high velocities from
near the neutron star into the helium and even hydrogen shells of the star.
In this section, the early postshock convection and its interaction with the
hydrodynamic instabilities at the composition interfaces of the progenitor star
will be discussed.

\subsection{Convective Overturn in the Neutrino-Heated Region}

Convective instabilities in the layers adjacent to the nascent neutron star
are a natural consequence of the negative entropy gradient built up by
the weakening of the prompt shock prior to its stagnation and by neutrino
heating~\cite{JKRref:bet90}. This was verified by two- and three-dimensional 
simulations~\cite{JKRref:hera92,JKRref:milwil,JKRref:heetal,JKRref:buetal,JKRref:heetal,JKRref:shimiz,JKRref:jamu95,JKRref:jamu96,JKRref:mez98b}. 
Figure~\ref{JKR:fig-4} shows the entropy distribution between
proto-neutron star and supernova shock about 300~ms after core bounce
for one such calculation~\cite{JKRref:kifon,JKRref:kietal}.
Although there is general agreement about the existence and the growth of
hydrodynamic instabilities in the layer between the shock at $R_{\rm s}$
and the radius of maximum neutrino heating (which is just outside the gain
radius, $R_{\rm g}$), the strength of the convective overturn
and its importance for the success of the neutrino-heating mechanism
are still a matter of debate.

Two-dimensional simulations with a spectrally averaged, flux-limited 
diffusion treatment of neutrino transport~\cite{JKRref:heetal,JKRref:buetal},
or with the neutrino luminosity being given as a free parameter at the inner 
boundary, which replaces the neutron star at the 
center~\cite{JKRref:jamu95,JKRref:jamu96}, 
found successful explosions in cases where spherically symmetric models
fail (Fig.~\ref{JKR:fig-3}).
According to these simulations, the convective overturn in the
neutrino-heated region has the following effects on the shock propagation.
Heated matter from the region close to
the gain radius rises outward and at the same time is exchanged with cool
gas flowing down from the shock. Since the production reactions
of neutrinos ($e^{\pm}$ capture on nucleons and thermal processes) are
very temperature sensitive, the expansion and cooling of rising plasma
reduces the energy loss by the reemission of neutrinos. Moreover, the net energy
deposition by neutrinos is enhanced as more cool material is exposed to the
large neutrino fluxes near the gain radius 
(the radial dilution of the fluxes goes roughly as $1/r^2$).
Since hot matter moves towards the shock, the pressure behind the shock 
increases, an effect which pushes the shock farther out. This leads
to a growth of the gain region and therefore also of the net energy transfer
from neutrinos to the stellar gas, favoring an explosion.

The consequences of postshock convection are clearly visible from the 
results plotted in Fig.~\ref{JKR:fig-3}, where the 
explosion energy $E_{>0}$ as a function of time is shown for spherically
symmetric and two-dimensional 
calculations of the same post-collapse model, but with different
assumed neutrino luminosities from the proto-neutron 
star~\cite{JKRref:jamu95,JKRref:jamu96}.
$E_{>0}$ is defined to include the sum of internal, kinetic,
and gravitational energy for all zones where this sum is positive
(the gravitational binding energies of stellar mantle and envelope
and additional energy release from nuclear burning are not taken into
account). For one-dimensional simulations with $\nu_e$ luminosities 
(and very similar $\bar\nu_e$ luminosities) below
$1.9\cdot 10^{52}\,$erg/s explosions could not be obtained when the 
proto-neutron star was assumed static, and the threshold value of 
the luminosity was $2.2\cdot 10^{52}\,$erg/s when the neutron star was 
contracting. The supporting effects of convective overturn
between the gain radius and the shock lead to explosions even below the
critical value in spherical symmetry, and to a faster development of the 
explosion. 

Simulations with a better description of the neutrino transport
by a multi-energy-group treatment of the neutrino diffusion~\cite{JKRref:mez98b},
confirm the existence of such convective processes in the region of neutrino 
heating, but the associated effects are not strong enough to revive the
stalled prompt supernova shock, although the outward motion of the
shock is enhanced. 

Fully self-consistent, multi-dimensional calculations, however, have not 
yet been done with a state-of-the-art Boltzmann neutrino transport, which
has recently become applicable for spherically symmetric models (see Section~3).
The current multi-dimensional simulations therefore demonstrate only the 
presence and potential importance of convection, but final conclusions on 
the viability of the neutrino-heating mechanism in the presence of
postshock convection are not possible at the moment. A quantitatively
meaningful description of the shock
revival phase, however, requires an accurate description of the transport
as well as a multi-dimensional approach.

%
%
\begin{figure}[t]
\begin{center}
\includegraphics[width=.95\textwidth]{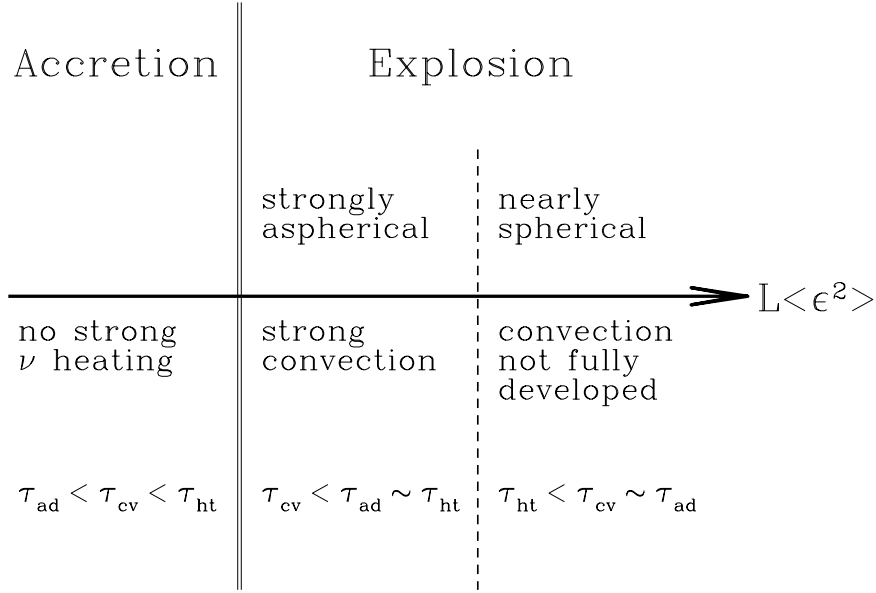}
\end{center}
\caption[]{
Order scheme for the post-collapse dynamics in dependence of
$L_{\nu}\langle\epsilon_{\nu}^2\rangle$, which determines the strength of
the neutrino heating outside of the neutrinosphere. The destiny of the star 
--- accretion or explosion --- can be understood by the relative size of the
timescales of neutrino heating, $\tau_{\rm ht}$, matter advection through the
gain region onto the nascent neutron star, $\tau_{\rm adv}$, and growth
of convective instabilities, $\tau_{\rm cv}$.
}
\label{JKR:fig-5}
\end{figure}


\subsection{Is Neutrino-Driven Convection Crucial for an Explosion?}

The role of convective overturn for the development of an explosion becomes
clearer by considering the three timescales of neutrino heating,
$\tau_{\rm ht}$, advection of accreted matter through the gain layer
into the cooling region and down to the neutron star (compare Fig.~\ref{JKR:fig-1}),
$\tau_{\rm ad}$, and the timescale for the growth of convective instabilities, 
$\tau_{\rm cv}$.
The evolution of the shock --- accretion or explosion --- is determined by the
relative size of these three timescales. Straightforward considerations
show that they are of the same order and the destiny of the star is therefore
a result of a tight competition between the different processes.

The heating timescale is estimated from the initial entropy per nucleon,
$s_{\rm i}$, the critical entropy $s_{\rm c}$ (Eq.~(\ref{JKR:eq-4})),
and the heating rate per nucleon (Eq.~(\ref{JKR:eq-1})) as
\begin{equation}
\tau_{\rm ht}
\,\approx\,{s_{\rm c}-s_{\rm i}\over Q_{\nu}^+/(k_{\rm B}T)}
\,\approx\,45\,{\rm ms}\cdot {s_{\rm c}-s_{\rm i}\over 5 k_{\rm B}/N}
\,{R_{{\rm g},7}^2(T/2{\rm MeV})\,f\over(L_{\nu}/4\cdot 10^{52}{\rm erg/s})
\langle\epsilon_{\nu,15}^2\rangle} \; ,
\label{JKR:eq-7}
\end{equation}
for $L_{\nu}$ being the total luminosity of $\nu_e$ plus $\bar\nu_e$.
With a postshock velocity of
$u_1 = u_0/\beta \approx (\gamma-1)\sqrt{GM/R_{\rm s}}/(\gamma+1)$
the advection timescale is
\begin{equation}
\tau_{\rm ad}\,\approx\,
{R_{\rm s}-R_{\rm g}\over u_1}\,\approx\,55\,{\rm ms}\cdot
\left(1-{R_{\rm g}\over R_{\rm s}}\right)\,{R_{{\rm s},200}^{3/2}\over
\sqrt{M_1}} \; ,
\label{JKR:eq-8}
\end{equation}
where the gain radius can be determined as
\begin{equation}
R_{{\rm g},7}\,\cong \,0.9\, T_{\mathrm{s}}^{3/2} R_{\mathrm{s},200}^{3/2}
\,f^{1/4}
\left( {L_{\nu}\over 4\cdot 10^{52}{\rm erg/s}}\right)^{-1/4}
\langle\epsilon_{\nu,15}^2\rangle^{-1/4} 
\label{JKR:eq-9}
\end{equation}
from the requirement that the heating rate, Eq.~(\ref{JKR:eq-1}), is equal to
the cooling rate per nucleon,
$Q_{\nu}^-\approx 2.3\, T^6\,{\rm MeV}/(N\cdot {\rm s})$, when
$R_{\mathrm{s},200}$ is the shock radius in units of 200~km, 
$Y_n + Y_p \approx 1$ is assumed, and 
use is made of the power-law behavior of the temperature according to
$T(r)\approx T_{\mathrm{s}}(R_{\mathrm{s}}/r)$, with
$T_{\mathrm{s}}$ being the postshock temperature in MeV.
The growth timescale
of convective instabilities in the neutrino-heated region depends on the
gradients of entropy and lepton number through the growth rate of
Ledoux convection, $\sigma_{\rm L}$:
\begin{eqnarray}
\tau_{\rm cv}\,\approx\,{\ln{(100)}\over \sigma_{\rm L}} &\approx&
4.6\left\lbrace{g\over\rho}\left\lbrack
\left({\partial\rho\over\partial s}\right)_{\! Y_e,P}
{{\rm d}s\over{\rm d}r}+\left({\partial\rho\over\partial Y_e}\right)_{\! s,P}
{{\rm d}Y_e\over{\rm d}r}
\right\rbrack\right\rbrace^{-1/2}\nonumber \\
&\sim& 20\,{\mathrm{ms}} \cdot 
\left({R_{\mathrm{s}}\over R_{\mathrm{g}}}-1\right)^{1/2} 
{R_{\mathrm{g},7}^{3/2}\over \sqrt{M_1}} \ .
\label{JKR:eq-10}
\end{eqnarray}
The numerical value was obtained with the gravitational acceleration
$g = GM/R_{\mathrm{g}}^2$, 
$(\partial\rho/\partial s)_P \sim -\rho/s$, and 
${\mathrm{d}}s/{\mathrm{d}}r \sim - {1\over 2}s/(R_{\mathrm{s}}-R_{\mathrm{g}})$.
The term proportional to the gradient of $Y_e$ was assumed to be negligible.
$\tau_{\rm cv}$ of Eq.~(\ref{JKR:eq-10})
is sensitive to the detailed conditions between 
gain radius (close to which $s$ develops a maximum)
and the shock. The neutrino heating timescale is shorter for larger values
of the neutrino luminosity $L_{\nu}$ and mean squared neutrino energy
$\langle\epsilon_{\nu}^2\rangle$. All three timescales, $\tau_{\rm ht}$, $\tau_{\rm ad}$
and $\tau_{\mathrm{cv}}$, decrease roughly in the same way with smaller
gain radius or shock position.

In order to be a crucial help for the explosion, convective overturn in the
neutrino-heated region must develop on a sufficiently short timescale.
This happens only in a rather narrow window of
$L_{\nu}\langle\epsilon_{\nu}^2\rangle$ where
$\tau_{\rm cv}<\tau_{\rm ad}\sim\tau_{\rm ht}$ (Fig.~\ref{JKR:fig-5}). 
For smaller neutrino
luminosities the heating is too weak to create a sufficiently large entropy
maximum, and rapid convective motions cannot develop before the accreted gas is
advected through the gain radius ($\tau_{\rm ad}<\tau_{\rm cv}<\tau_{\rm ht}$).
In this case neither with nor without convective processes energetic explosions
can occur (Fig.~\ref{JKR:fig-3}). 
For larger neutrino luminosities the neutrino heating is so strong,
and the heating timescale correspondingly short
($\tau_{\rm ht}<\tau_{\rm cv}\sim\tau_{\rm ad}$), that expansion of the
postshock layers has set in before the convective activity reaches a
significant level. In this case convective overturn is an unavoidable
side-effect of the neutrino heating behind the shock, but is not 
necessary for starting the explosion.

The parametric studies performed by Janka and 
M\"uller~\cite{JKRref:jamu95,JKRref:jamu96}
support this discussion, which helps one understanding the seemingly 
discrepant results obtained by different groups.

%
%
\begin{figure}[t!]
\begin{center}
\vspace{3truecm}
\end{center}
\caption[]{
Snapshot from a two-dimensional simulation of the explosion of
a blue supergiant star with 15 solar masses at a time 1170 seconds after
the stellar core has collapsed to a neutron star~\cite{JKRref:kietal}. 
In the left half of the
figure the density is shown in a region with a radius of about 2.2 million
kilometers, in the right half three color images of the mass densities
of radioactive nickel (red and pink), silicon (green, light blue, whitish)
and oxygen (deep blue) are superposed. One can see that the ejecta of the
explosion are
inhomogeneous and anisotropic, and the original onion-shell structure of the
exploding star was shredded. Nickel is concentrated in dense, fast-moving
clumps along the extended filaments seen in the left plot.
}
\label{JKR:fig-4a}
\end{figure}
%
%

%
%
\begin{figure}[t]
\begin{center}
\includegraphics[width=.475\textwidth]{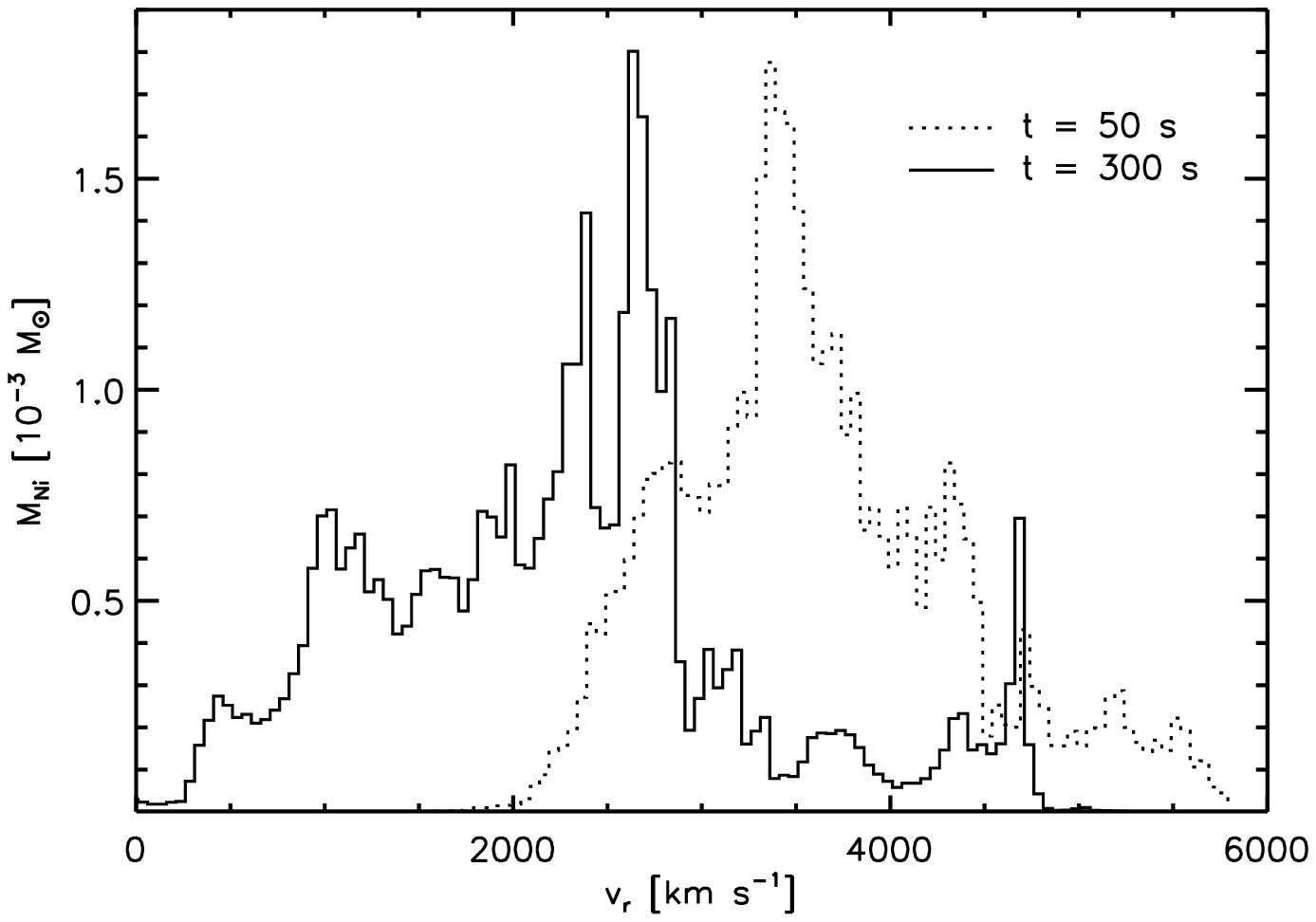}
\hspace{5pt}
\includegraphics[width=.475\textwidth]{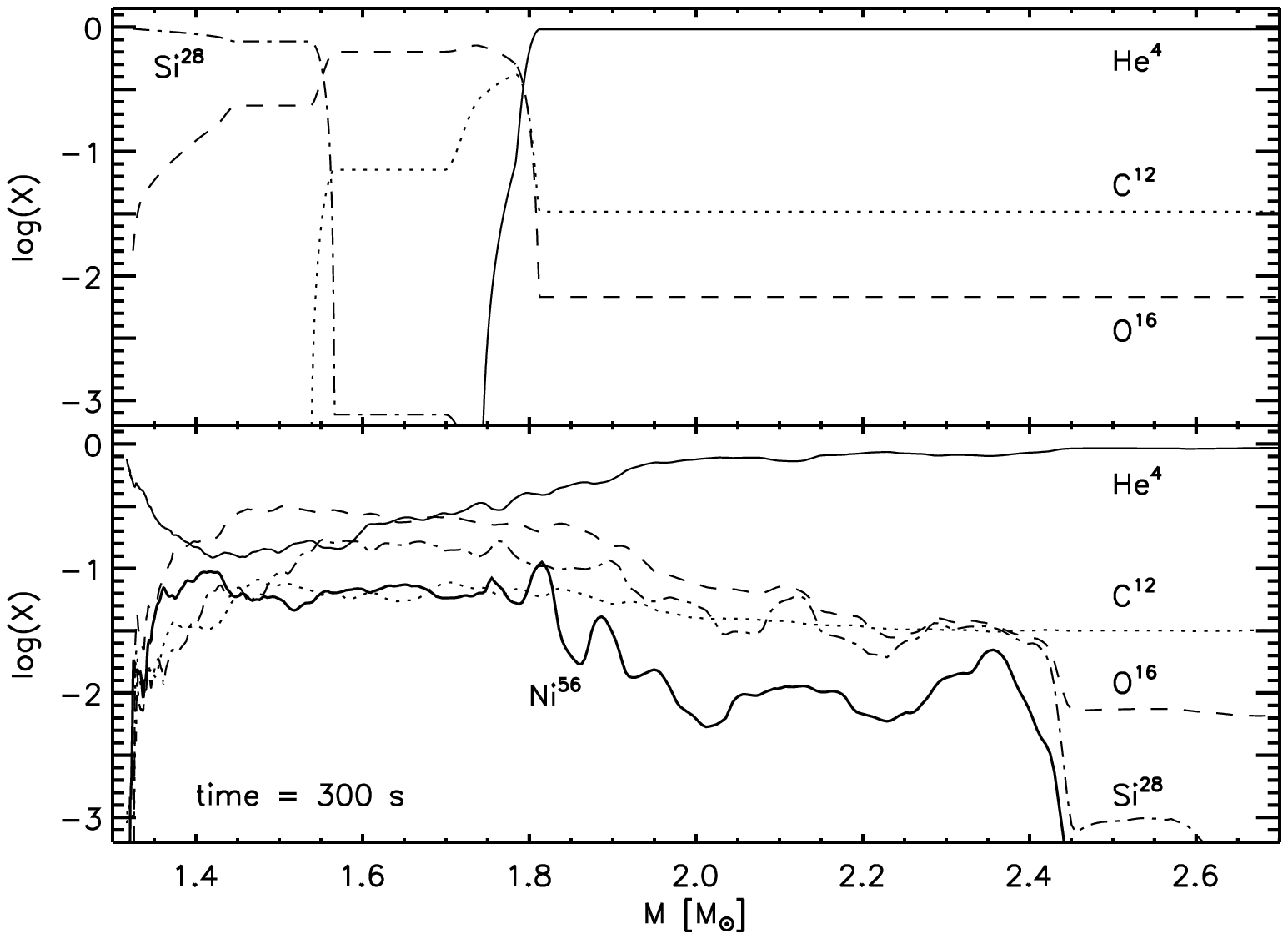}
\end{center}
\caption[]{
{\em Left:} Distribution of $^{56}$Ni vs.\ radial velocity at $t = 50$~s
and $t = 300$~s after core bounce.
{\em Right:} Initial composition of the star exterior to the iron core (top)
and composition 300~s after core bounce (bottom). C, O, Si, and the newly 
synthesized Ni have been mixed beyond the inner half of the helium core,
and He has been carried inward.
}
\label{JKR:fig-4b}
\end{figure}


\subsection{Nucleosynthesis and Mixing Instabilities}

Besides increasing the efficiency of neutrino energy deposition, convection in
the postshock layer has an important influence also on the nucleosynthesis and
distribution of radioactive elements. In particular, nickel is not produced
by silicon burning 
in a spherical shell, but is concentrated in dense clumps and pockets between
rising bubbles of neutrino-heated matter in the expanding postshock layer
(Fig.~\ref{JKR:fig-4}). 

The further evolution of the shock until it breaks out of the stellar
surface hours later, was recently followed by using 
adaptive mesh refinement techniques~\cite{JKRref:kifon,JKRref:kietal}. 
These allow for a dynamic
adjustment of the computational grid such that small structures can be 
treated with high resolution, while the whole computation covers a huge volume. 

A few seconds after its formation, the shock has passed the silicon and oxygen 
layers and 
propagates through the helium shell of the star. The initial anisotropies
have been compressed into a narrow shell, from which new instabilities start
to grow. Rayleigh-Taylor mushrooms penetrate into the helium layer and carry
O, Si and Ni farther out, while He sinks in. Within minutes, long,
dense filaments reach far into the helium shell, associated with them 
fast-moving knots that contain dominant contributions of different heavy
elements from the deeper layers (Fig.~\ref{JKR:fig-4a}). Nickel, silicon and 
oxygen move through helium with velocities up to several 1000~km/s 
(Fig.~\ref{JKR:fig-4b}). 

These simulations show that the
hydrodynamic instabilities which occur in the first second of the 
explosion, do not only play a role during the phases of shock rejuvenation and
nickel formation. They act also as seed perturbations for the
instabilities at the composition interfaces of the progenitor, which finally
destroy the onion-shell structure of the pre-collapse star. 

The results are in good agreement with observations of mixing and
anisotropies in many Type Ib,c supernovae. In case of Supernova~1987A, 
a Type II explosion of a massive star which has retained its hydrogen
envelope, the observed high nickel velocities in the hydrogen envelope
cannot be explained by the models. The nickel clumps are strongly
decelerated at the He/H interface, where they enter a dense helium ``wall''
which builds up after the passage of the shock. The dissipation of the
kinetic energy of the clumps does not allow nickel to penetrate into the 
hydrogen layer with high velocities.

\section{Neutron Star Formation}

Convective energy transport inside the newly formed neutron star can increase
the neutrino luminosities considerably~\cite{JKRref:burlat}. This can be
crucial for energizing the stalled supernova 
shock~\cite{JKRref:wima88,JKRref:wima93}. 

Convection in the neutron
star can be driven by gradients of the entropy and/or proton
(electron lepton number) fraction in the nuclear
medium~\cite{JKRref:epst}. The type of instability
which grows most rapidly, e.g., doubly diffusive neutron-finger
convection~\cite{JKRref:wima88,JKRref:wima93} or
Ledoux convection~\cite{JKRref:burlat} or
quasi-Ledoux convection~\cite{JKRref:keil,JKRref:kejamu}, may be 
a matter of the properties of the nuclear equation of state, which
determines the magnitudes and signs of the thermodynamic
derivatives~\cite{JKRref:brudi}. It is also sensitive to the gradients
that develop, and thus may depend on
the details of the treatment of neutrino transport in the dense interior
of the star.

Convection below the neutrinosphere
seems to be disfavored during the very early post-bounce evolution by the
currently most elaborate supernova
models~\cite{JKRref:brume,JKRref:brmedi,JKRref:mez98a}, but can develop
deeper inside the nascent neutron star on a longer timescale
($\ga 100\,$ms after bounce) and can encompass the whole star within
seconds~\cite{JKRref:burlat,JKRref:kejamu,JKRref:keil}.

\begin{figure}[t]
\begin{center}
\includegraphics[width=.475\textwidth]{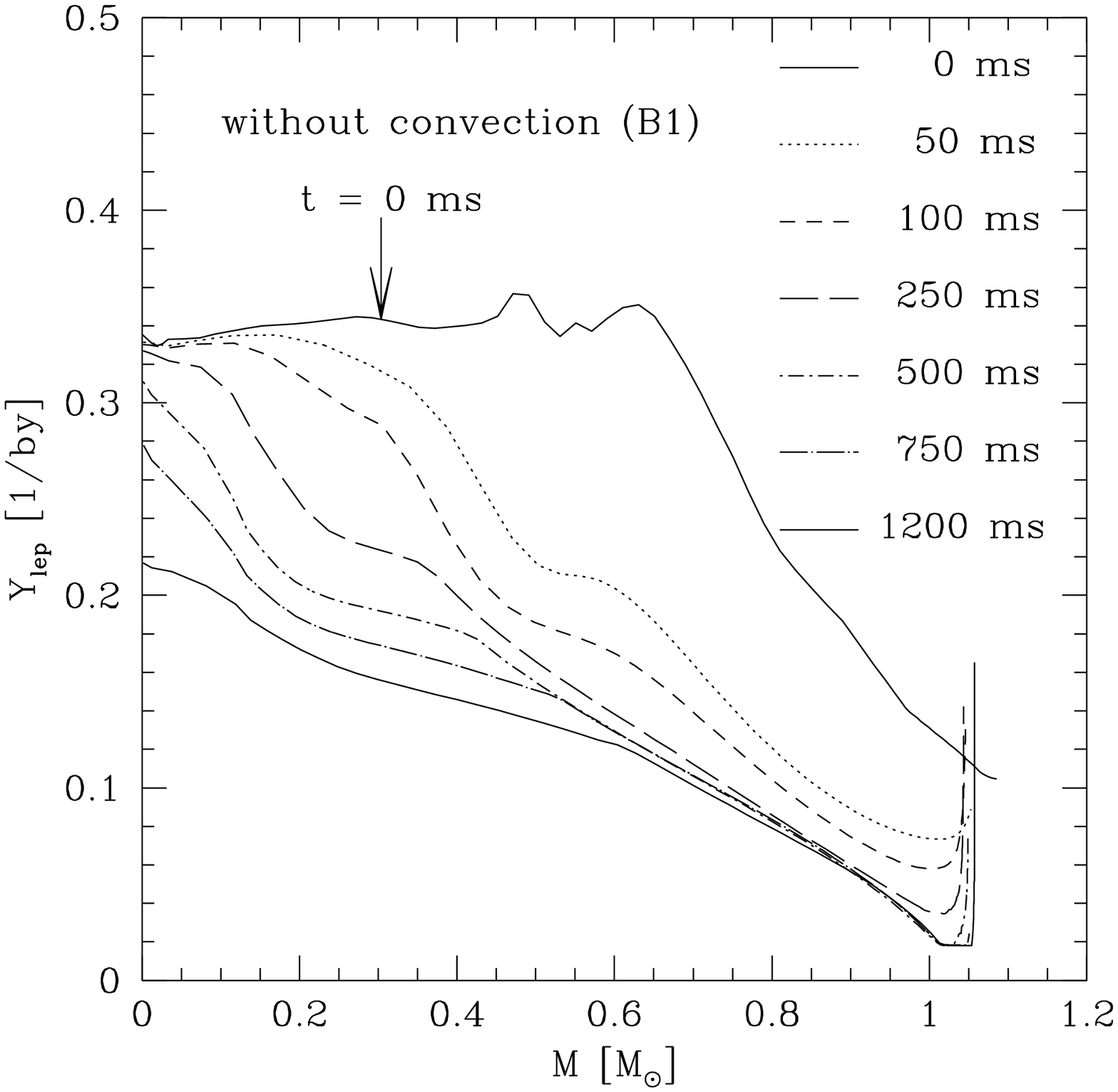}
\hspace{5pt}
\includegraphics[width=.475\textwidth]{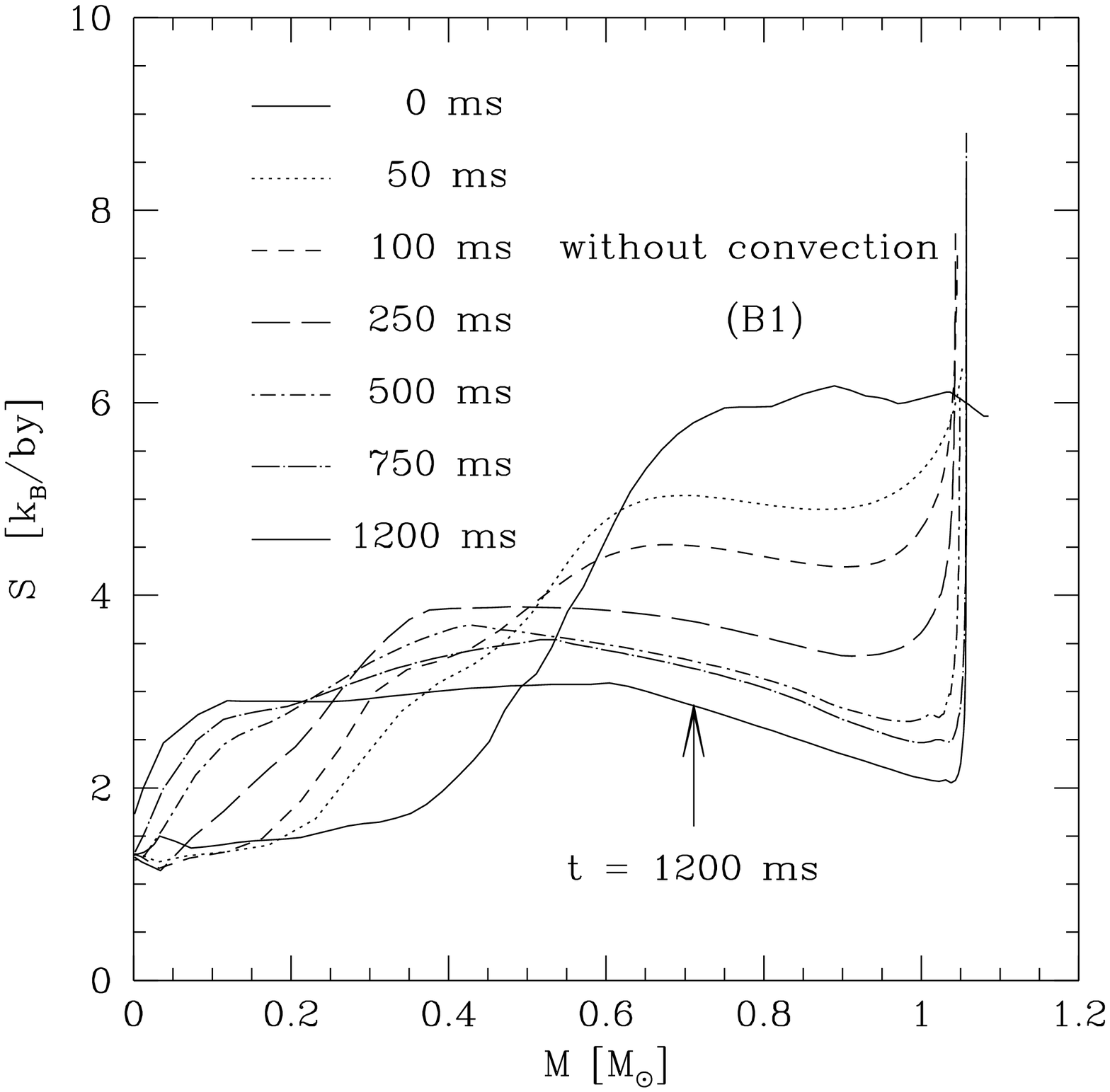}
\end{center}
\caption[]{
Profiles of the lepton fraction $Y_{\rm lep} = n_{\rm lep}/n_b$
(left) and of the entropy per nucleon, $s$, (right) as functions of enclosed
(baryonic) mass for different times in a one-dimensional simulation of
the neutrino cooling of a $\sim 1.1\,M_{\odot}$ proto-neutron star.
Negative gradients of lepton number and entropy suggest potentially
convectively unstable regions. Time is (roughly) measured from core bounce.
}
\label{JKR:fig-5a}
\end{figure}

Negative lepton number and entropy gradients have been seen in several 
one-dimensional (spherically symmetric) simulations of the neutrino-cooling 
phase of nascent neutron 
stars~\cite{JKRref:bula86,JKRref:burlat,JKRref:keja,JKRref:sumi} 
(see also Fig.~\ref{JKR:fig-5a}) 
and have suggested the existence of regions which are potentially unstable 
against Ledoux convection. 
Recent calculations~\cite{JKRref:pons} with improved neutrino opacities
of the nuclear medium, which were described consistently with
the employed equation of state, confirm principal
aspects of previous simulations, in particular the existence of
Ledoux-unstable layers in the neutron star.

\subsection{Convection inside the Nascent Neutron Star}

Two-dimensional, hydrodynamical simulations
were performed for the neutrino-cooling phase of a $\sim 1.1\,M_{\odot}$
proto-neutron star that formed in the core collapse of a $15\,M_{\odot}$ 
star~\cite{JKRref:kejamu,JKRref:keil}.
The models followed the evolution for a period of
more than 1.2~seconds. They demonstrate the development of 
convection and its importance for the cooling and deleptonization of the 
neutron star. 

The simulations were carried out with the hydrodynamics
code {\it Prometheus}. A general relativistic 1D gravitational potential 
with Newtonian corrections for asphericities was used, 
$\Phi\equiv\Phi_{\rm 1D}^{\rm GR}+(\Phi_{\rm 2D}^{\rm N}-\Phi_{\rm 1D}^{\rm N})$,
and a flux-limited (equilibrium) neutrino diffusion scheme was applied
for each angular bin separately (``1${1\over 2}$D'').

%
%


%
%
\begin{figure}[t]
\begin{center}
\includegraphics[width=.475\textwidth]{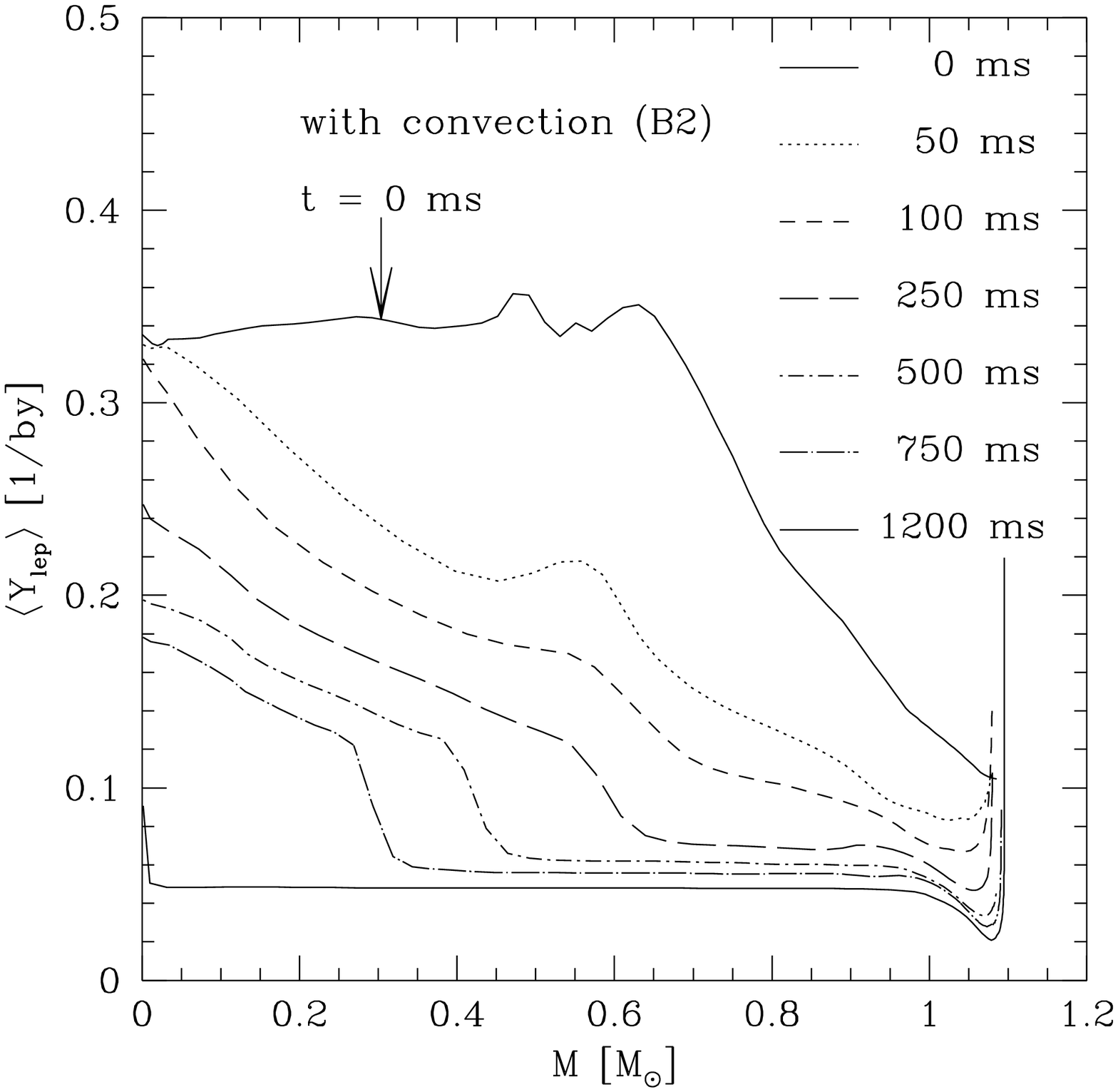}
\hspace{5pt}
\includegraphics[width=.475\textwidth]{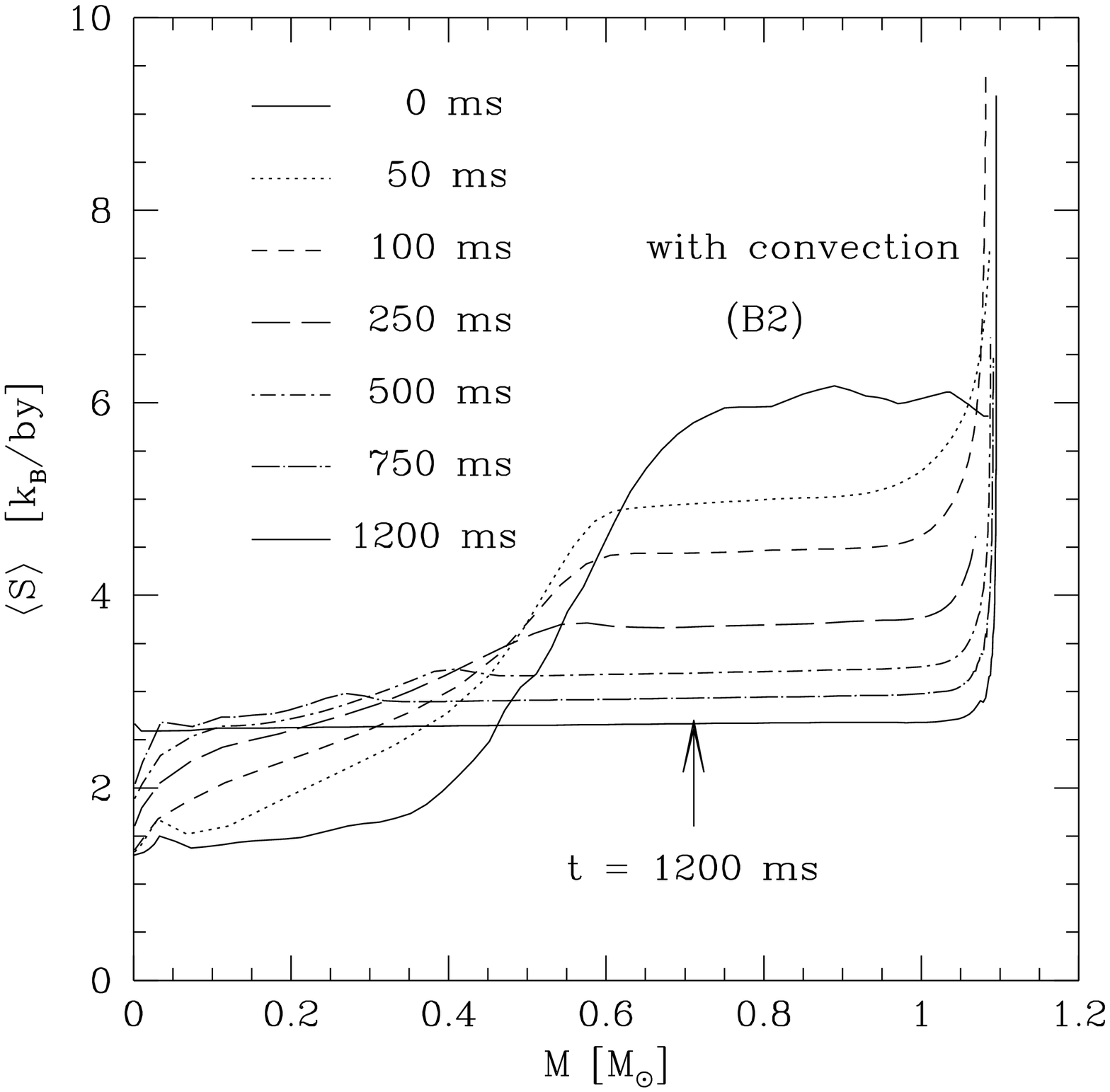}
\end{center}
\caption[]{
Same as Fig.~\ref{JKR:fig-5a}, but for a two-dimensional, hydrodynamical
simulation which allowed to follow the development of convection. The plots
show angularly averaged quantities in the $\sim 1.1\,M_{\odot}$ proto-neutron
star. In regions with convective activity the gradients of $Y_{\rm lep}$ and
$s$ are flattened. The convective layer encompasses an increasingly larger
part of the star.
}
\label{JKR:fig-6}
\end{figure}

%

%
%
\begin{figure}[t]
\begin{center}
\vspace{3truecm}
\end{center}
\caption[]{
Absolute values of the convective velocity in the proto-neutron star
for two instants (about $0.5\,{\rm s}$ (left) and $1\,{\rm s}$ (right)
after core bounce) as obtained in a two-dimensional, hydrodynamical
simulation. The arrows indicate the direction of the velocity field. Note that
the neutron star has contracted from a radius of about 60~km initially to
little more than 20~km. The growth of the convective region can be seen.
Typical velocities of the convective motions are several $10^8\,$cm/s.
}
\label{JKR:fig-7}
\end{figure}

%

The simulations showed that convectively unstable surface-near regions 
(i.e., around the neutrinosphere and below an initial density of about
$10^{12}\,$g/cm$^3$) exist only for a short period of a few ten 
milliseconds after bounce, in agreement with the findings by other
groups~\cite{JKRref:brume,JKRref:brmedi,JKRref:mez98a}.
Due to a flat entropy profile and a negative lepton number gradient,
convection, however, also starts in a layer deeper inside the star, 
between an enclosed mass of $0.7\,M_{\odot}$ and $0.9\,M_{\odot}$, at
densities above several $10^{12}\,$g/cm$^3$. From there the convective 
region digs into the star and reaches the center after about one second
(Figs.~\ref{JKR:fig-6}, \ref{JKR:fig-7}, and \ref{JKR:fig-9}).
Convective velocities as high as $5\cdot 10^8\,$cm/s were found
(about 10--20\% of the local sound speed), corresponding to kinetic 
energies of up to 1--$2\cdot 10^{50}\,$erg (Fig.~\ref{JKR:fig-7}). 
Because of these high velocities and rather flat entropy and composition 
profiles in the star (Fig.~\ref{JKR:fig-6}), the overshooting region is
large (see Fig.~\ref{JKR:fig-9}). The same is true for undershooting during
the first $\sim 100\,$ms after bounce. Sound waves and perturbations are
generated in the layers above and interior to the convection zone.

The coherence lengths of convective structures are of the order of 20--40
degrees (in 2D!) (see Fig.~\ref{JKR:fig-7}) and coherence times are around
10~ms, which corresponds to only one or two overturns. The convective
pattern is therefore very time-dependent and nonstationary.
Convective motions lead to considerable variations of the composition.
The lepton fraction (and thus the abundance of protons) shows relative 
fluctuations of several 10\%. The entropy differences in rising and sinking
convective bubbles are much smaller, only a few per cent, while temperature
and density fluctuations are typically less than one per cent. 

\begin{figure}[t]
\begin{center}
\includegraphics[width=.65\textwidth]{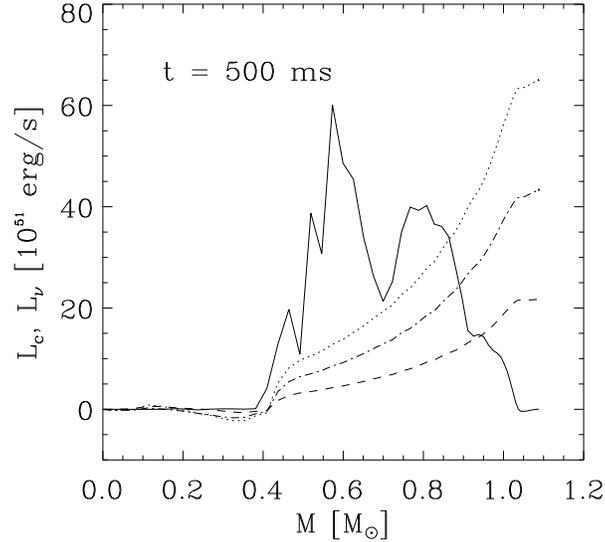}
\end{center}
\caption[]{
Convective ``luminosity'' (solid line) and neutrino luminosities (dashed:
$L_{\nu_e}+L_{\bar\nu_e}$, dash-dotted: $L_{\nu_{\mu}}+L_{\bar\nu_{\mu}}
+L_{\nu_{\tau}}+L_{\bar\nu_{\tau}}$, dotted: total) as functions of
enclosed baryonic mass for the two-dimensional proto-neutron star
simulation about 500~ms after core bounce.
}
\label{JKR:fig-8}
\end{figure}

The energy transport in the neutron star is dominated by neutrino diffusion
near the center, whereas convective transport plays the major role in a thick 
intermediate layer where the convective activity is strongest. Radiative
transport takes over again when the neutrino mean free path becomes large
near the surface of the star (Fig.~\ref{JKR:fig-8}). 
But even in the convective layer the convective
energy flux is only a few times larger than the diffusive flux. This means
that neutrino diffusion is not negligibly small in the convective region.
This fact has important consequences for the driving mechanism
of the convection.

%
%
\begin{figure}[t]
\begin{center}
\vspace{3truecm}
\end{center}
\caption[]{
{\em Left:} Convectively unstable region (corresponding to negative
values of the displayed quantity $\omega_{\rm QL}^2 = -(g/\rho)C_{\rm QL}$
with $C_{\rm QL}$ from Eq.~(\ref{JKR:eq-12})) about 500~ms after bounce
according to the Quasi-Ledoux criterion which includes non-adiabatic and
lepton-transport effects by neutrino diffusion.
{\em Right:} Layer of Quasi-Ledoux convective instability (blue)
as function of time for a two-dimensional simulation.
The angle-averaged criterion $C_{\rm QL}^{\rm 1D}(r)\equiv
\min_{\theta}\left(C_{\rm QL}(r,\theta)\right) > 0$
with $C_{\rm QL}(r,\theta)$ from Eq.~(\ref{JKR:eq-12}) is plotted.
The dotted area is outside of the computed
star, green denotes stable layers where over- and undershooting causes
lateral velocities with angularly averaged absolute values of
$\ave{|v_{\theta}|} > 10^7\,{\rm cm\,s}^{-1}$, and white are
convectively ``quiet'' regions of the star.
}
\label{JKR:fig-9}
\end{figure}

%

\subsection{Driving Force of Convection}

The convective
activity in the neutron star cannot be explained by, and considered as 
ideal Ledoux convection. Applying the Ledoux criterion for local instability,
\begin{equation}
C_{\rm L}(r,\theta)\,=\,{\rho\over g}\,\sigma_{\rm L}^2\,=\, 
\left({\partial\rho\over\partial s}\right)_{\! Y_{\rm lep},P}
{{\rm d}s\over{\rm d}r}+
\left({\partial\rho\over\partial Y_{\rm lep}}\right)_{\! s,P}
{{\rm d}Y_{\rm lep}\over{\rm d}r} \,>\,0 \; ,
\label{JKR:eq-11}
\end{equation}
with $\sigma_{\rm L}$ from
Eq.~(\ref{JKR:eq-10}) and $Y_e$ replaced by the total lepton fraction 
$Y_{\rm lep}$ in the neutrino-opaque interior of the neutron star
(for reasons of simplicity, $\nabla s$ was replaced by ${\rm d}s/{\rm d}r$
and $\nabla Y_{\rm lep}$ by ${\rm d}Y_{\rm lep}/{\rm d}r$),
one finds that the convecting region should actually be stable, despite
of slightly negative entropy {\it and} lepton number gradients. In fact, 
below a critical value of the lepton fraction (e.g., 
$Y_{\rm lep,c} = 0.148$ for $\rho = 10^{13}\,$g/cm$^3$ and $T = 10.7\,$MeV)
the thermodynamical derivative $(\partial \rho/\partial Y_{\rm lep})_{s,P}$
changes sign and becomes positive because of nuclear and Coulomb forces in
the high-density equation of state~\cite{JKRref:brudi}.
Therefore negative lepton number gradients
should stabilize against convection in this regime. However, an idealized
assumption of Ledoux convection is not fulfilled in the situations considered
here: Because of neutrino diffusion, energy exchange and, in particular,
lepton number exchange between convective elements and their surroundings 
are {\it not} negligible. Taking the neutrino transport effects on 
$Y_{\rm lep}$ into account in a modified 
``{\it Quasi-Ledoux criterion}''~\cite{JKRref:keil}, 
\begin{equation}
C_{\rm QL}(r,\theta)\equiv
\left({\partial\rho\over\partial s}\right)_{\! 
\langle Y_{\rm lep}\rangle,\langle P\rangle}
{{\rm d}\langle s\rangle \over{\rm d}r}+
\left({\partial\rho\over\partial Y_{\rm lep}}\right)_{\! 
\langle s\rangle,\langle P\rangle}\!
\left({{\rm d}\langle Y_{\rm lep}\rangle \over{\rm d}r} 
- \beta_{\rm lep} {{\rm d}Y_{\rm lep}\over {\rm d}r}\right)
>0 \, ,
\label{JKR:eq-12}
\end{equation}
one determines instability exactly where
the two-dimensional simulation reveals convective activity. 
In Eq.~(\ref{JKR:eq-12}) the quantities $\langle Y_{\rm lep}\rangle$
and $\langle s\rangle$ mean averages over the polar angles $\theta$,
and local gradients have to be distinguished from gradients of 
angle-averaged quantities which describe the stellar background. The
term $\beta_{\rm lep} ({\rm d}Y_{\rm lep}/{\rm d}r)$ with 
the empirically determined value $\beta_{\rm lep}\approx 1$ accounts 
for the change of the lepton concentration along the path of a rising
fluid element due to neutrino diffusion. Figure~\ref{JKR:fig-9} shows that 
about half a second 
after core bounce strong, driving forces for convection occur
in a narrow ring between 9 and 10~km, where a steep negative gradient
of the lepton fraction exists (see Fig.~\ref{JKR:fig-6}). Farther out,
convective instability is detected only in finger-like structures of
rising, high-$Y_{\rm lep}$ gas.

%
%
\begin{figure}[t]
\begin{center}
\vspace{3truecm}
\end{center}
\caption[]{
Absolute value of the gas velocity in a convecting,
rotating proto-neutron star about $750\,$ms after bounce (left). Convection
is suppressed near the rotation axis (vertical) and develops strongly only
near the equatorial plane where a flat distribution of the specific
angular momentum $j_z$ (right) has formed.
}
\label{JKR:fig-10}
\end{figure}

%

\subsection{Accretion and Rotation}

In other two-dimensional models, post-bounce mass accretion and rotation of the
forming neutron star were included. Accretion causes stronger convection
with larger velocities in a more extended region. This
can be explained by the steepening of lepton number and entropy gradients
and the increase of the gravitational potential energy when additional 
matter is added onto the neutron star. 

Rotation has very interesting consequences,
e.g., leads to a suppression of convective motions near the rotation
axis because of a stabilizing stratification of the specific angular
momentum (see Fig.~\ref{JKR:fig-10}), an effect which can be understood by 
applying the (first) Solberg-H\o iland criterion for instabilities in
rotating, self-gravitating bodies~\cite{JKRref:tass}:
\begin{equation}
C_{\rm SH}(r,\theta)\,\equiv\,
{1\over x^3}\,{{\rm d}j_z^2\over {\rm d}x} + 
{{\vec a}\over \rho}\left\lbrack 
\left({\partial\rho\over\partial s}\right)_{\! Y_{\rm lep},P}
\nabla s +
\left({\partial\rho\over\partial Y_{\rm lep}}\right)_{\! s,P}
\nabla Y_{\rm lep} \right\rbrack
\,<\,0  \; .
\label{JKR:eq-13}
\end{equation}
Here, $j_z$ is the specific angular momentum of a fluid element, which is 
conserved for axially symmetric configurations, $x$ is the distance 
from the rotation axis, and in case of rotational
equilibrium ${\vec a}$ is the sum of gravitational and centrifugal
accelerations, ${\vec a} = \nabla P/\rho$. Changes of the lepton number
in rising or sinking convective elements due to neutrino diffusion were
neglected in Eq.~(\ref{JKR:eq-13}). Ledoux (or Quasi-Ledoux) convection
can only develop where the first term is not too positive. 
In Fig.~\ref{JKR:fig-10} fully developed convective motion is therefore 
constrained to a zone of nearly constant $j_z$ close to the equatorial
plane. At higher latitude the convective velocities are much smaller, and
narrow, elongated convective cells aligned with cylindrical regions of 
$j_z = {\rm const}$ parallel to the rotation axis are visible.

The rotation pattern displayed in Fig.~\ref{JKR:fig-10} is highly differential 
with a rotation period of $7.3\,$ms at $x = 22\,$km and of $1.6\,$ms at
$x = 0.6\,$km. It has self-consistently developed under the influence of
neutrino transport and convection when the neutron star had contracted
from an initial radius of about $60\,$km (with a surface rotation period
of $55\,$ms at the equator and a rotation period of $\sim 5\,$ms near the 
center) to a final radius
of approximately $22\,$km. Due to the differential nature of the rotation,
the ratio of rotational kinetic energy to the gravitational potential
energy of the star is only 0.78\% in the beginning and a few per cent at
the end after about $1\,$s of evolution.

%
%
\begin{figure}[t]
\begin{center}
\includegraphics[width=.475\textwidth]{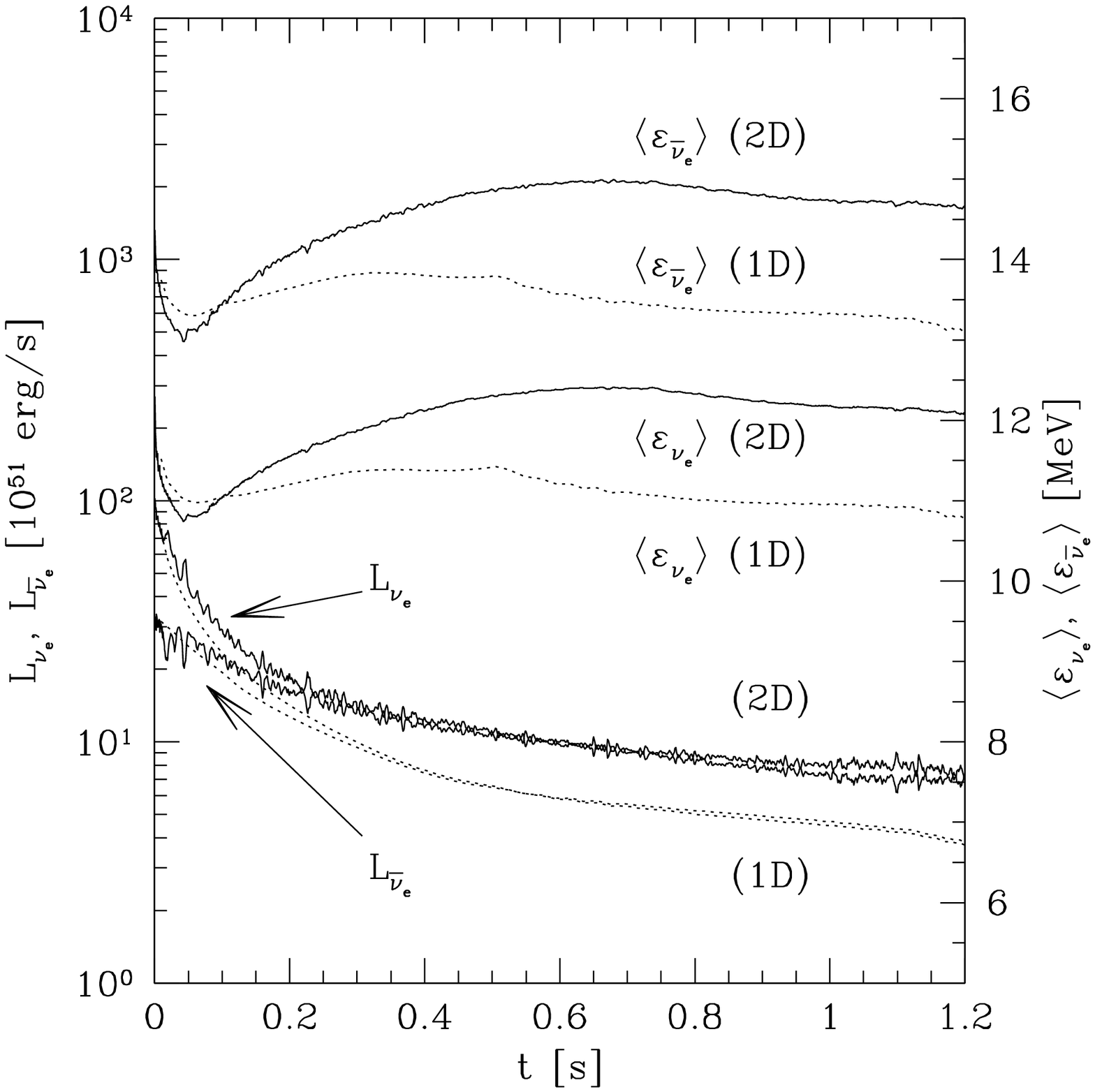}
\hspace{5pt}
\includegraphics[width=.475\textwidth]{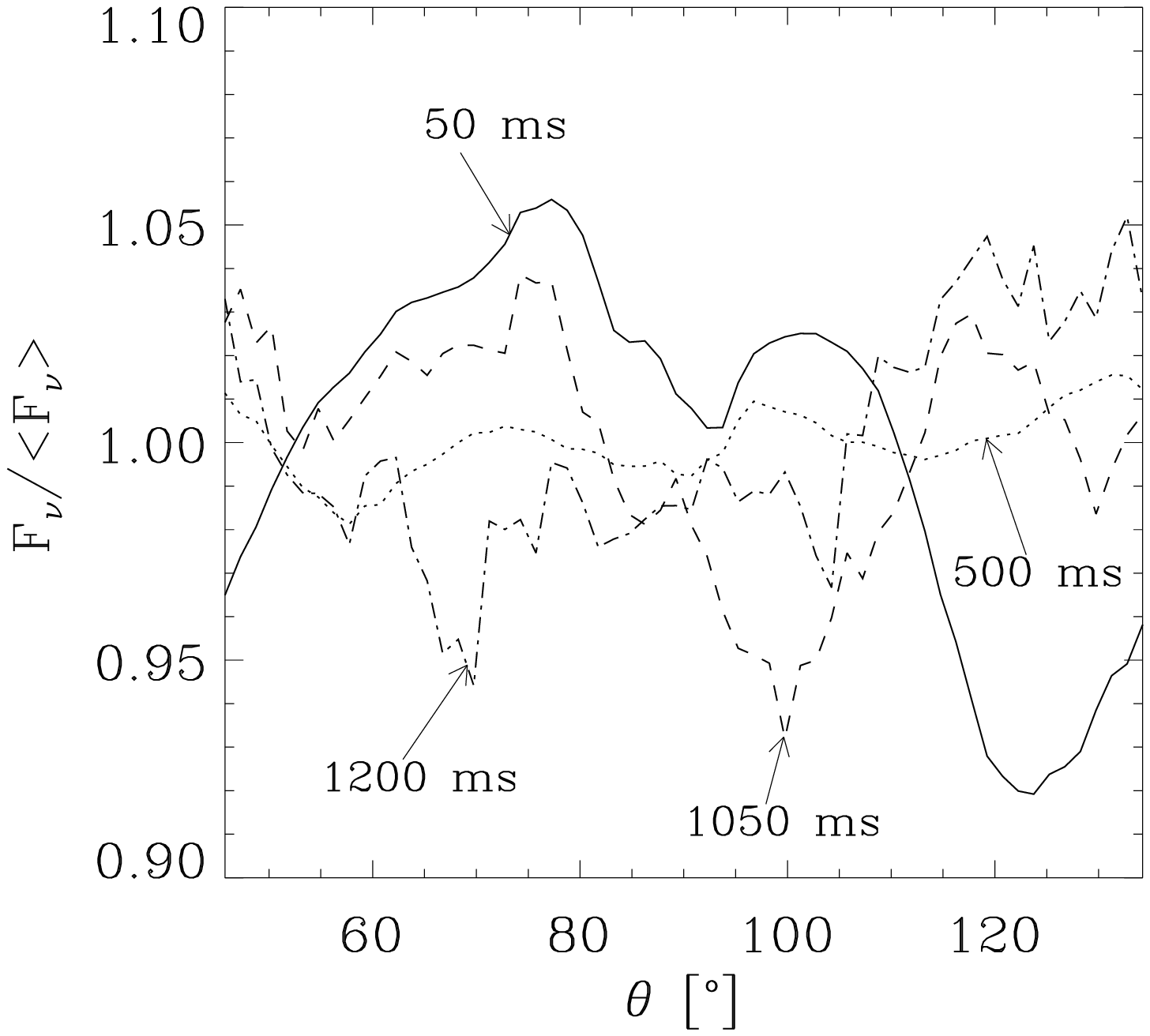}
\end{center}
\caption[]{
{\em Left:} Luminosities $L_{\nu}(t)$ and mean energies
$\langle\epsilon_{\nu}\rangle(t)$ of $\nu_e$ and $\bar\nu_e$ for a
$1.1\,M_{\odot}$ proto-neutron star without (``1D''; dotted) and with convection
(``2D''; solid).
{\em Right:} Angular variations of the neutrino flux at
different times for the 2D simulation.
}
\label{JKR:fig-11}
\end{figure}

%

%
%
\begin{figure}[t]
\begin{center}
\includegraphics[width=.65\textwidth]{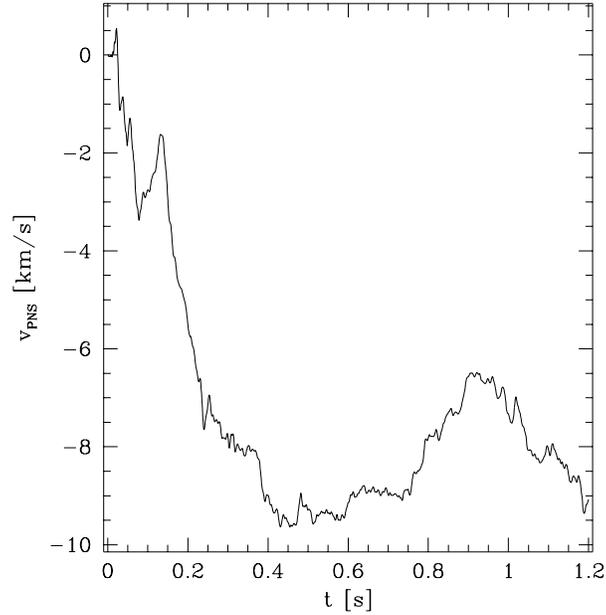}
\end{center}
\caption[]{
Kick velocity of the neutron star as a function of time, caused by the 
anisotropic emission of neutrinos due to convection. The two-dimensional
simulation was done with a polar grid from 0 to $\pi$.
}
\label{JKR:fig-12}
\end{figure}

\subsection{Consequences of Proto-Neutron Star Convection}

Convection inside the proto-neutron star can raise the neutrino luminosities 
within a few hundred ms after core bounce (Fig.~\ref{JKR:fig-11}).
In the considered collapsed core of a $15\,M_{\odot}$ star, $L_{\nu_e}$ and
$L_{\bar\nu_e}$ increase by up to 50\% and the mean neutrino energies by 
about 15\% at times later than 200--300~ms post bounce.
This favors neutrino-driven explosions on timescales of a few
hundred milliseconds after shock formation. Also, the deleptonization of
the nascent neutron star is strongly accelerated, raising the $\nu_e$ 
luminosities relative to the $\bar\nu_e$ luminosities during this time. 
This helps to increase the electron fraction $Y_e$ in the neutrino-heated
ejecta and might solve the overproduction problem of $N=50$ nuclei 
during the early epochs of the explosion~\cite{JKRref:kejamu}.
In case of rotation, the effects of convection on
the neutrino emission depend on the direction. Since strong convection 
occurs only close to the equatorial plane, the neutrino fluxes are
convectively enhanced there, while they are essentially unchanged near the 
poles. 

Anisotropic mass motions due to convection in the 
neutron star lead to gravitational wave emission and anisotropic radiation
of neutrinos. The angular variations of the neutrino flux found in the 
2D simulations are of the order of 5--10\% (Fig.~\ref{JKR:fig-11}). With the
typical size of the convective cells and the
short coherence times of the convective structures, the 
global anisotropy of the neutrino emission from the cooling proto-neutron
star is very small. This implies a kick velocity of the nascent neutron star
due to anisotropic neutrino emission of only $\sim 10$~km/s in a 2D simulation
(Fig.~\ref{JKR:fig-12}).
Because the convective elements are likely to become even smaller in 3D,
kick velocities of 300~km/s or even more, as observed for many pulsars, 
can definitely not be explained by convectively perturbed neutrino emission.

%
%
\begin{figure}[t]
\begin{center}
\includegraphics[width=.475\textwidth]{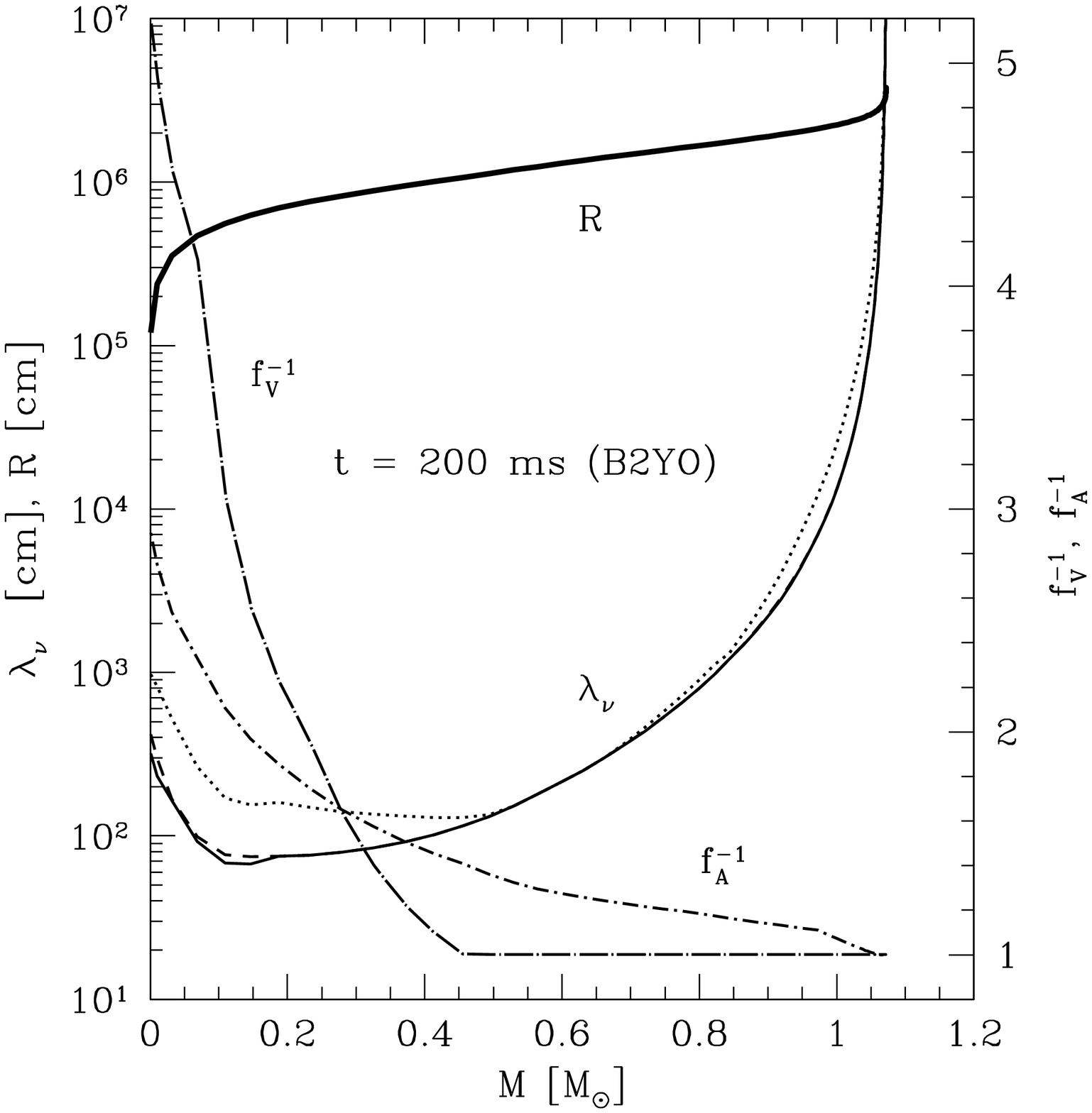}
\hspace{5pt}
\includegraphics[width=.475\textwidth]{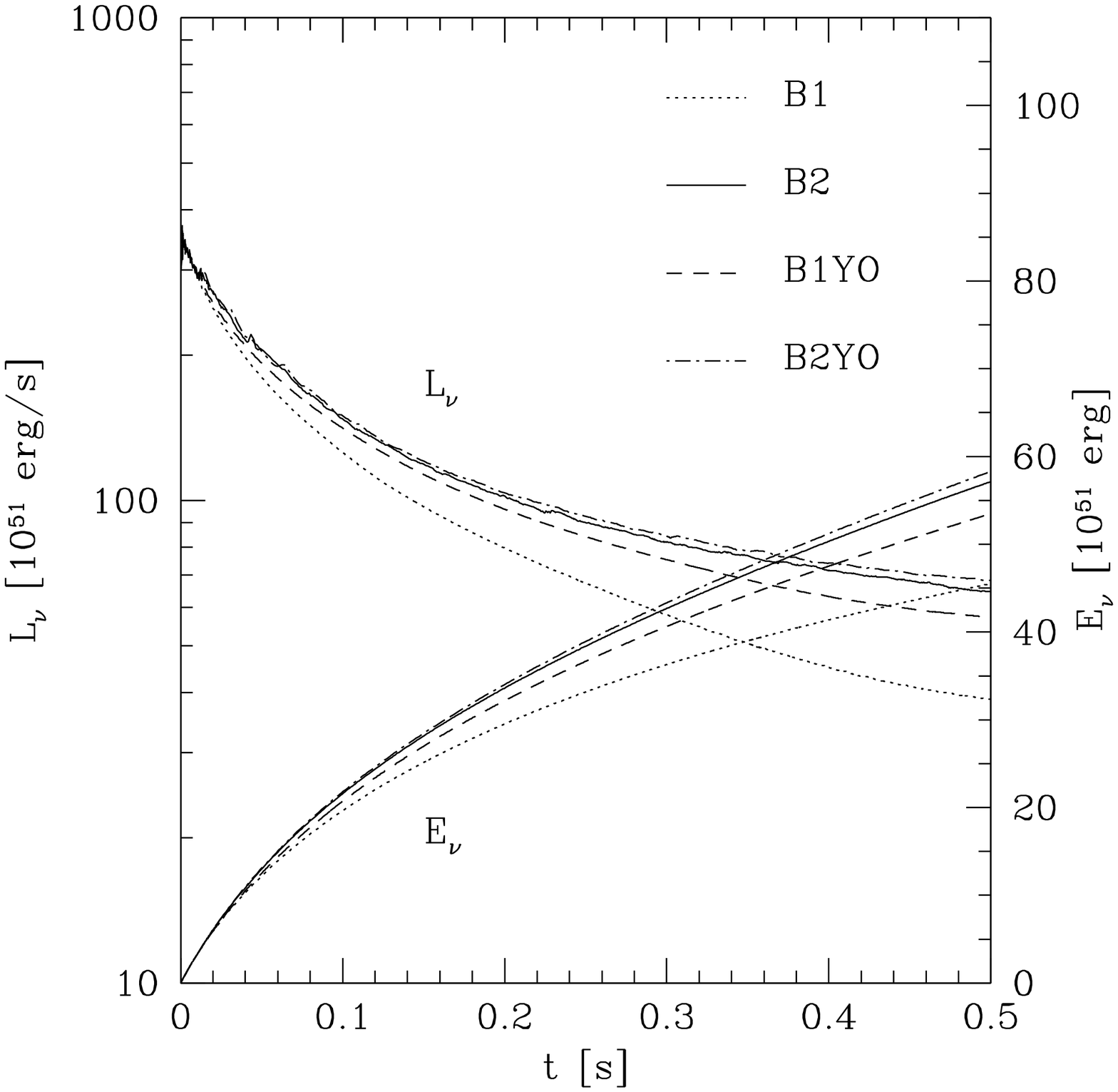}
\end{center}
\caption[]{
{\it Left:} Thermal averages of the neutrino mean free paths for
$\nu_e$ absorption (dotted line), $\nu_e$ scattering
(solid line), and muon and tau neutrino scattering
(dashed line), respectively, according to the standard description
of the neutrino opacities. $M$ is the baryonic mass enclosed by
the radial coordinate $R$ (bold solid line) of the newly formed neutron
star about $200\,{\rm ms}$ after core bounce. Also shown are the factors
$f_{\rm A}^{-1}$ and $f_{\rm V}^{-1}$ (dash-dotted lines) which give
a measure of the increase of the neutrino mean free paths caused by a
suppression of the axial-vector and vector current contributions to the
neutrino opacities due to in-medium effects~\cite{JKRref:yamada}
(Yamada, personal communication).
{\it Right:} Total neutrino luminosities, $L_{\nu}$, and integrated
energy loss, $E_{\nu}$, as functions of time for spherically symmetric
models without convection (B1 and B1YO) and two-dimensional models with
convection (B2 and B2YO). Models B1 and B2 were computed with standard
neutrino opacities whereas in B1YO and B2YO in-medium suppression of
the neutrino opacities was included~\cite{JKRref:yamada}
(Yamada, personal communication).
}
\label{JKR:fig-13}
\end{figure}


\subsection{Neutrino Opacities in Nuclear Matter and Neutron Star Convection}

Another important issue of interest are the neutrino opacities in
the dense and hot nuclear medium of the nascent neutron star. In
current supernova models, the description of neutrino-nucleon 
interactions is incomplete because the standard approximations
assume isolated and infinitely massive nucleons~\cite{JKRref.tub75}. 
Therefore effects like the fermion phase space blocking of the
nucleons, the reduction of the effective nucleon mass 
by momentum-dependent nuclear interactions in the dense plasma,
and nucleon thermal motions and recoil are either 
neglected completely or approximated in a more or less reliable
manner~\cite{JKRref:bru85,JKRref:bula86}. These effects have
been recognized to be 
important~\cite{JKRref:sch90,JKRref:jak96,JKRref:pra97,JKRref:red97}
for calculations of the neutrino luminosities and spectra,
but still await careful inclusion in supernova codes. For this
purpose a consistent description of nuclear equation of state and
neutrino-matter interactions is desirable.

Many-body (spatial) correlations due to strong
interactions~\cite{JKRref:saw89,JKRref:hor91,JKRref:bur98,JKRref:red97,JKRref:yamada}
and multiple-scattering effects by spin-dependent forces between 
nucleons (temporal spin-density 
correlations)~\cite{JKRref:raf95,JKRref:han97}
are of particular interest, because they lead to
a reduction of the neutrino opacities in the newly formed neutron star 
and are associated with additional modes of energy transfer between
neutrinos and the nuclear medium. 

A reduction of the neutrino opacities implies
larger neutrino mean free paths and thus increases the neutrino luminosities.
(Fig.~\ref{JKR:fig-13} and Refs.~\cite{JKRref:kei95,JKRref:pons,JKRref:bur98}).
The neutrino diffusion is accelerated most strongly in the very dense core
of the nascent neutron star. Convection in the intermediate region between
core and outer layers turns out not to be suppressed, but is still the
fastest mode of energy
transport. Therefore reduced neutrino opacities as well as convective
energy transport are important, but the combined effects do not 
appreciably change the convectively enhanced neutrino emission
(Fig.~\ref{JKR:fig-13}, right)~\cite{JKRref:jak01}.

\section{Summary}

Supernova explosions of massive stars are an important phenomenon for 
applying nuclear and particle physics, in particular neutrino
physics. The processes going on in the extremely dense
and hot core of the exploding star are accessible to direct 
measurements only through neutrinos or gravitational waves.
Empirical information about the events that cause the explosion 
and accompany the formation of a neutron star, however, can also be 
deduced from observable characteristics of supernovae, for example
their explosion energy or the amount and distribution of radioactive nuclei,
and from the properties of neutron stars.

Theoretical models need to establish the link between the core
physics and these observables. In the past years it has been recognized that
hydrodynamical instabilities and mixing processes on large scales play
an important role within the core as well as in the outer layers of the
exploding star. Convection can change the cooling of the nascent neutron 
star, supports the revival of the stalled shock by neutrino heating, and 
destroys the onion-shell structure of the progenitor star. Multi-dimensional
calculations are therefore necessary to understand why and how supernovae 
explode, and to make predictions for their observable consequences.

Spherically symmetric simulations, Newtonian and general relativistic, 
with the most advanced treatment of neutrino
transport by solving the Boltzmann equation, do not produce explosions.  
This emphasizes the importance of convection, but may also point to
physics still missing in the models. One such weakness of current 
simulations is an overly simplified description of neutrino interactions
with nucleons in the nuclear medium of the neutron star. A kinematically
correct treatment of these reactions, taking into account nucleon thermal 
motions, recoil and fermi blocking, needs only a technical step, but a
better understanding of the effects of nucleon correlations and their
consistent treatment with the equation of state requires theoretical
progress. 

The neutrino-heating mechanism, although the favored explanation for
the explosion, is still controversial, both because of the status of 
modeling and because of observations which seem hard to explain. 
Although significant progress has
been made, multi-dimensional simulations with an accurate and reliable
handling of neutrino transport and an up-to-date treatment of the input 
physics are still missing, and definite conclusions can therefore not 
be drawn at the moment.

\bigskip\noindent
{\bf Acknowledgements.}
H.-Th. Janka thanks E. M\"uller and W. Keil for many years
of fruitful and enjoyable collaboration. This work was supported by
the Son\-der\-for\-schungs\-be\-reich 375 on ``Astroparticle Physics'' of the 
Deut\-sche For\-schungs\-ge\-mein\-schaft.

\end{document}